\documentclass[intlimits,twoside,a4paper]{article}

\usepackage{amsmath,amssymb}
\usepackage{graphicx}
\usepackage{color}

\usepackage[T2A]{fontenc}
\usepackage[cp1251]{inputenc}

\usepackage{multirow}

\usepackage[eqsecnum]{cmpj2}

\issue{2015}{18}{3}{32601}
\doinumber{10.5488/CMP.18.32601}

\articletype{Review}

\title[Molecular theory of solvation]
{Molecular theory of solvation: Methodology summary and illustrations\footnote{This paper is a part of the collection of papers published in Condens. Matter
 Phys., 2015, \textbf{18}, No~1.}}

\author{A.~Kovalenko\refaddr{addr1,addr2}}

\authorcopyright{A.~Kovalenko, 2015}
\addresses{
\addr{addr1} National Institute for Nanotechnology, 11421 Saskatchewan Dr., Edmonton, AB, T6G 2M9, Canada
\addr{addr2} Department of Mechanical Engineering, University of Alberta, Mechanical Engineering Bldg. 4--9, Edmonton, AB, T6G 2G7, Canada
}

\date{Received March 16, 2015, in final form June 22, 2015}
\begin{document}

\maketitle

\begin{abstract}
Integral equation theory of molecular liquids based on statistical mechanics  is quite promising as an essential part of multiscale methodology for chemical and biomolecular nanosystems in solution. Beginning with a molecular interaction potential force field, it uses diagrammatic analysis of the solvation free energy to derive integral equations for correlation functions between molecules in solution in the statistical-mechanical ensemble. The infinite chain of coupled integral equations for many-body correlation functions is reduced to a tractable form for 2- or 3-body correlations by applying the so-called closure relations. Solving these equations produces the solvation structure with accuracy comparable to molecular simulations that have converged but has a critical advantage of readily treating the effects and processes spanning over a large space and slow time scales, by far not feasible for explicit solvent molecular simulations. One of the versions of this formalism, the three-dimensional reference interaction site model (3D-RISM) integral equation complemented with the Kovalenko-Hirata (KH) closure approximation, yields the solvation structure in terms of 3D maps of correlation functions, including density distributions, of solvent interaction sites around a solute (supra)molecule with full consistent account for the effects of chemical functionalities of all species in the solution. The solvation free energy and the subsequent thermodynamics are then obtained at once as a simple integral of the 3D correlation functions by performing thermodynamic integration analytically.
%
%
Analytical form of the free energy functional permits the self-consistent field coupling of 3D-RISM-KH with quantum chemistry methods in multiscale description of electronic structure in solution, the use of 3D maps of potentials of mean force as scoring functions for molecular recognition and protein-ligand binding in docking protocols for fragment based drug design, and the hybrid MD simulation running quasidynamics of biomolecules steered with 3D-RISM-KH mean solvation forces. The 3D-RISM-KH theory has been validated on both simple and complex associating liquids with different chemical functionalities in a wide range of thermodynamic conditions, at different solid-liquid interfaces, in soft matter, and various environments and confinements. The 3D-RISM-KH theory offers a ``mental microscope'' capable of providing an insight into structure and molecular mechanisms of formation and functioning of various chemical and biomolecular systems and nanomaterials.
\pacs  61.20.-p, 61.20.Gy, 65.20.-w, 81.16.Fg, 82.60.-s, 87.15.A-
\end{abstract}

\section{Introduction}
\label{sec:Introduction}

Nanoscale properties, phenomena, and processes are profoundly different from macroscopic laws governing the behavior of continuous media and materials. All functional features of nanostructures stem from the microscopic properties of constituting atoms but manifest at length scale from one to hundreds nanometers and time scale up to microseconds and more. By changing size, composition, and fabrication protocol, nanostructure properties and processes can be tuned in the widest range \cite{Feynmann:1960:23:22}. Therefore, predictive modelling of nanosystems should operate at length scales from an Angstr\"{o}m to hundreds nanometers and microns and time scales to milliseconds and seconds, and yet derive their properties from the chemical functionalities of the constituents. (An example is biological cellular systems of protein nanomachines operating in crowded environment.) Explicit molecular modelling of such nanosystems involves millions and billions of molecules and is by far not feasible in a ``brute force'' approach employing just {\it ab initio} quantum chemical methods and/or molecular simulations. Addressing this challenge requires the development and use of multiscale methods coupling several levels of description, from electronic structure methods for building blocks and classical molecular simulations for critical aggregates in the system, to statistical-mechanical theories for their large assemblies and mean properties in a statistical ensemble over characteristic size and time scales, to eventually come up with macroscopic scale properties of the nanostructures and related processes showing up in the ``real observable world''. A true, genuine challenge of multiscale modelling is the development of a theoretical framework that couples methods at different scales, so that observables at lower-level scales are analytically linked in a self-consistent field description to force fields of more coarse-grained models at higher-level scales \cite{RocoMirkinHersam:2011}. Statistical mechanics itself, in particular, integral equation theory of liquids \cite{HansenMcDonald:2006}, is an example of such a theoretical coupling between microscopic molecular variables and thermodynamic, macroscopic properties.

\section{Statistical-mechanical, molecular theory of solvation}
\label{sec:MolecularSolvationTheory}

Molecular theory of solvation, also known as reference interaction site model (RISM) \cite{HansenMcDonald:2006, Hirata:2003}, is based on the first principles foundation of statistical mechanics and Ornstein-Zernike (OZ) type integral equation theory of molecular liquids \cite{HansenMcDonald:2006}. It provides a firm platform to handle complex chemical and biomolecular systems in solution. As distinct from molecular simulations which explore the phase space of a molecular system by direct sampling, RISM theory operates with spatial distributions rather than trajectories of molecules and is based on analytical summation of the free energy diagrams which yields the solvation structure and thermodynamics in the statistical-mechanical ensemble. It yields a solvation structure by solving the RISM integral equations for correlation functions and then the solvation thermodynamics analytically as a single integral of a closed form in terms of the correlation functions obtained. Its three-dimensional (3D) version, 3D-RISM theory gives 3D maps of solvent distributions around a solute macromolecule of arbitrary shape \cite{Chandler:1986:85:5971, Chandler:1986:85:5977, Beglov:1995:101:7821, Kovalenko:1998:290:237, Kovalenko:1999:110:10095, Kovalenko:2000:112:10391, Kovalenko:2000:112:10403, Kovalenko:2003:169, Gusarov:2012:33:1478, Kovalenko:2013:85:159}. An important component enabling a utility of 3D-RISM theory for complex systems in solution has been the closure relation proposed by Kovalenko and Hirata (KH approximation) \cite{Kovalenko:1999:110:10095, Kovalenko:2003:169, Kovalenko:2013:85:159}. For simple and complex solvents and solutions of a given composition, including buffers, salts, polymers, ligands and other cofactors at a finite concentration, the 3D-RISM-KH molecular theory of solvation properly accounts for chemical functionalities by representing in a single formalism both electrostatic and non-polar features of solvation, such as hydrogen bonding, structural solvent molecules, salt bridges, solvophobicity, and other electrochemical, associative and steric effects. For real systems, to solve the 3D-RISM-KH integral equations is far less computationally expensive than to run molecular simulations which should be long enough to sample relevant exchange and binding events. This enables handling complex systems and processes occurring on large space and long time scales, which is problematic and frequently not feasible for molecular simulations. The 3D-RISM-KH theory has been validated on both simple and complex associating liquids and solutes with different chemical functionalities \cite{Kovalenko:2013:85:159, Gusarov:2006:110:6083, Casanova:2007:3:458, Miyata:2008:29:871, Kaminski:2010:114:6082, Malvaldi:2009:113:3536, Kovalenko:2012:8:1508, Kovalenko:2001:349:496, Kovalenko:2002:1:381, Yoshida:2002:106:5042, Omelyan:2003:2:193, Huang:2015:119:5588}, including ionic liquids \cite{Malvaldi:2009:113:3536}, and polyelectrolyte gels \cite{Kovalenko:2012:8:1508}, in a range of fluid thermodynamic conditions \cite{Hirata:2003, Kovalenko:2001:349:496, Kovalenko:2002:1:381, Yoshida:2002:106:5042, Omelyan:2003:2:193}, in various environments such as interfaces with metal \cite{Kovalenko:2000:112:10391, Kovalenko:2003:169}, metal oxide \cite{Shapovalov:2000:320:186}, zeolite \cite{Gusarov:2012:33:1478, Stoyanov:2011:203}, clay \cite{Fafard:2013:117:18556, Huang:2014:118:23821}, and in confinement of carbon nanotubes \cite{Casanova:2007:3:458}, synthetic organic rosette nanotubes \cite{Kovalenko:2013:85:159, Moralez:2005:127:8307, Johnson:2007:129:5735, Yamazaki:2010:11:361}, nanocellulose based bionanomaterials \cite{Silveira:2013:135:19048, Stoyanov:2014:29:144, Silveira:2015:6:206}, and biomolecular systems \cite{Kovalenko:2013:85:159, Luchko:2010:6:607, Omelyan:2013:39:25, Omelyan:2013:139:244106, Omelyan:2015:11:1875, Yoshida:2006:128:12042, Yoshida:2009:113:873, Imai:2009:131:12430, Phongphanphanee:2008:130:1540, Phongphanphanee:2010:132:9782, Maruyama:2011:3:290, Yonetani:2008:128:186102, Maruyama:2010:114:6464, Harano:2001:114:9506, Imai:2001:59:512, Imai:2002:112:9469, Imai:2007:16:1927, Kovalenko:2015:22:575, Blinov:2010:98:282, Yamazaki:2008:95:4540, Blinov:2011:37:718, Genheden:2010:114:8505, Kovalenko:2011:164:101, Nikolic:2012:8:3356, Imai:2011:115:8288, Huang:2015:55:317, Stumpe:2011:115:319}. The latter include case studies ranging from the structure of hydrated alanine dipeptide \cite{Luchko:2010:6:607, Omelyan:2013:39:25, Omelyan:2013:139:244106, Omelyan:2015:11:1875}, miniprotein 1L2Y and protein G \cite{Omelyan:2015:11:1875}; structural water, xenon, and ions bound to lysozyme protein \cite{Yoshida:2006:128:12042, Yoshida:2009:113:873}; selective binding and permeation of water, ions and protons in channels \cite{Kovalenko:2013:85:159, Kovalenko:2012:8:1508, Yoshida:2009:113:873, Imai:2009:131:12430, Phongphanphanee:2008:130:1540, Phongphanphanee:2010:132:9782, Maruyama:2011:3:290}; salt-induced conformational transitions of DNA \cite{Maruyama:2011:3:290, Yonetani:2008:128:186102, Maruyama:2010:114:6464}; partial molar volume and pressure induced conformational transitions of biomolecules \cite{Harano:2001:114:9506, Imai:2001:59:512, Imai:2002:112:9469, Imai:2007:16:1927, Kovalenko:2015:22:575}; formation and conformational stability of $\beta$-sheet Amyloid-$\beta$ (A$\beta$) oligomers \cite{Blinov:2010:98:282}, HET-s Prion and A$\beta$ fibrillous aggregates \cite{Yamazaki:2008:95:4540, Blinov:2011:37:718}; ligand-binding affinities of seven biotin analogues to avidin in aqueous solution \cite{Genheden:2010:114:8505}; binding modes of inhibitors of pathologic conversion and aggregation of prion proteins \cite{Kovalenko:2012:8:1508, Blinov:2011:37:718}, binding modes of thiamine against extracytoplasmic thiamine binding lipoprotein MG289 \cite{Kovalenko:2011:164:101, Nikolic:2012:8:3356}, binding modes and ligand efflux pathway in multidrug transporter AcrB \cite{Imai:2011:115:8288}, and maltotriose ligand binding to maltose-binding protein in which structural water triggers binding modes with protein domain motion resulting in the holo-closed structure that binds the ligand \cite{Huang:2015:55:317}; to aqueous electrolyte solution at physiological concentrations in biomolecular systems as large as the Gloebacter violaceus pentameric ligand-gated ion channel (GLIC) homologue in a lipid bilayer \cite{Kovalenko:2013:85:159, Kovalenko:2012:8:1508, Kovalenko:2011:164:101} and structural water promoting folding in the GroEL/ES chaperonin complex \cite{Stumpe:2011:115:319}. The 3D-RISM-KH theory provided an insight into such soft matter phenomena as the structure and stability of gels formed by oligomeric polyelectrolyte gelators in different solvents \cite{Kovalenko:2012:8:1508}.

\subsection{3D-RISM integral equations for molecular liquid structure}
\label{sec:3D-RISM}

The solvation structure is represented by the probability density $\rho_\gamma g_\gamma(\textbf{r})$ of finding interaction site $\gamma$ of solvent molecules at 3D space position $\textbf{r}$ around the solute molecule (which can be both a macromolecule and supramolecule), as given by the average number density $\rho_\gamma$ in the solution bulk times the normalized density distribution, or 3D distribution function, $g_\gamma(\textbf{r})$. The values of $g_\gamma(\textbf{r})>1$ and $g_\gamma(\textbf{r})<1$ indicate the areas of density enhancement or depletion, respectively, relative to the average density at a distance from the solute in the solution bulk where $g_\gamma \to 1$. The 3D distribution functions of solvent interaction sites around the solute molecule are determined from the 3D-RISM integral equation \cite{Chandler:1986:85:5971, Chandler:1986:85:5977, Beglov:1995:101:7821, Kovalenko:1998:290:237, Kovalenko:1999:110:10095, Kovalenko:2000:112:10391, Kovalenko:2000:112:10403, Kovalenko:2003:169, Gusarov:2012:33:1478, Kovalenko:2013:85:159}
\begin{equation}
  h_\gamma(\textbf{r}) = \sum_\alpha \int\textrm{d}\textbf{r}^\prime c_\alpha(\textbf{r}-\textbf{r}^\prime) \chi_{\alpha\gamma}(r^\prime) \, ,
\label{eq:3D-RISM}
\end{equation}
where $h_\gamma(\textbf{r})$ and $c_\gamma(\textbf{r})$ are, respectively, the 3D total and direct correlation functions of solvent site $\gamma$ around the solute molecule; $\chi_{\alpha\gamma}(r)$ is the site-site susceptibility of solvent (comprising both the intra- and intermolecular terms) which is an input to the 3D-RISM theory; and indices $\alpha$ and $\gamma$ enumerate all interaction sites on all sorts of solvent species. The diagrammatic analysis relates the density distribution function to the total correlation function as \cite{HansenMcDonald:2006}
\begin{equation}
  g_\gamma(\textbf{r}) = h_\gamma(\textbf{r}) + 1 \, ,
\label{eq:TCF}
\end{equation}
and so the latter has the meaning of the normalized probability density of 3D spatial correlations between the solute and solvent molecules, or normalized deviations of solvent site density around the solute molecule from its average value in the solution bulk. As follows from the diagrammatic expansion of the direct correlation function \cite{HansenMcDonald:2006}, the leading term of its asymptotics beyond the short-range region $D^{\textrm{sr}}$ of the solute-solvent repulsive core and attractive well (first solvation shell maximum),
\begin{equation}
  c_\gamma(\textbf{r}) \sim - u_\gamma(\textbf{r}) / (k_{\textrm{B}}T)   ~~~{\textrm{for}}~~~ \textbf{r} \notin D^{\textrm{sr}} \, ,
\label{eq:DCF-as}
\end{equation}
is given by the 3D interaction potential $u_\gamma(\textbf{r})$ between the whole solute molecule and solvent interaction site $\gamma$, scaled by the Boltzmann factor $k_{\textrm{B}}$ times temperature $T$. Inside the repulsive core, the direct correlation function strongly deviates from the asymptotics \eqref{eq:DCF-as} and assumes the values related to the solvation free energy of the solute molecule immersed in the solvent.

\subsection{KH closure approximation}
\label{sec:KH}

To obtain correlation functions, the 3D-RISM integral equation (1) should be complemented with another relation between the two functions $h_\gamma(\textbf{r})$ and $c_\gamma(\textbf{r})$ which is called a closure and which also involves the 3D solute-solvent site interaction potential $u_\gamma(\textbf{r})$ specified with a molecular force field. The exact closure has a nonlocal functional form which can be expressed as an infinite diagrammatic series in terms of multiple integrals of the total correlation function \cite{HansenMcDonald:2006}. However, these diagrams are cumbersome and the series suffers from a usual drawback of poor convergence, which makes useless to truncate the series at reasonably low-order diagrams and thus renders the exact closure computationally intractable. Therefore, it is replaced in practice with amenable approximations having analytical features that properly represent physical characteristics of the system, such as the long-range asymptotics of the correlation functions related to electrostatic forces and their short-range features related to the solvation structure and thermodynamics. Examples of closure relations to Ornstein-Zernike type integral equations (including RISM) suitable for liquids of polar and charged species are the so-called hypernetted chain (HNC) closure and the mean spherical approximation (MSA) \cite{HansenMcDonald:2006}, since they both enforce the long-range asymptotics like \eqref{eq:DCF-as}, which is critical for systems with charges. Such other well-known approximations as the Percus-Yevick, Modified Verlet, Martynov-Sarkisov, and Ballone-Pastore-Galli-Gazzillo closures reproduce solvation features for species with short-range repulsion but do not properly account for electrostatics in the interaction potential. By generalization, the 3D-HNC and 3D-MSA closures to the 3D-RISM integral equation \eqref{eq:3D-RISM} are constructed \cite{Chandler:1986:85:5971, Chandler:1986:85:5977, Beglov:1995:101:7821, Kovalenko:1998:290:237, Kovalenko:1999:110:10095, Kovalenko:2003:169}. While enforcing the asymptotics \eqref{eq:DCF-as}, the 3D-HNC closure strongly overestimates the association effects and, therefore, the 3D-RISM-HNC equations diverge for macromolecules with considerable site charges solvated in polar solvents or electrolyte solutions, which is almost always the case for biomolecules in aqueous solution. The 3D-MSA closure is free from that shortcoming but suffers from another serious deficiency of producing nonphysical negative values of the distribution function in the areas adjacent to the associative peaks. Those drawbacks are overcome with the approximation proposed by Kovalenko and Hirata (KH closure), the 3D version of which reads \cite{Kovalenko:1999:110:10095, Kovalenko:2003:169, Kovalenko:2013:85:159}
\begin{equation}
  g_\gamma(\textbf{r}) =
    \left\{
    \begin{array}{lcc}
      \exp\bigl[ - u_\gamma(\textbf{r}) / (k_{\textrm{B}}T) + h_\gamma(\textbf{r}) - c_\gamma(\textbf{r}) \bigr]
      & \textrm{for} & g_\gamma(\textbf{r}) \leqslant 1,
    \\ [6pt]
      1 - u_\gamma(\textbf{r}) / (k_{\textrm{B}}T) + h_\gamma(\textbf{r}) - c_\gamma(\textbf{r})
      & \textrm{for} & g_\gamma(\textbf{r}) > 1.
   \end{array} \right.
\label{eq:3D-KH}
\end{equation}
The 3D-KH closure relation \eqref{eq:3D-KH} couples in a nontrivial way the HNC approximation and the MSA linearization, the latter automatically applied to spatial regions of solvent density enhancement $g_\gamma(\textbf{r})>1$, including the repulsive core, and the latter to spatial regions of solvent density enrichment $g_\alpha(\textbf{r})>1$ such as strong peaks of association and long-range tails of near-critical fluid phases, and the former to spatial regions of density depletion $g_\alpha(\textbf{r})<1$ (including the repulsive core), while keeping the right asymptotics \eqref{eq:DCF-as} peculiar in both HNC and MSA. The distribution function and its first derivative are continuous at the joint boundary $g_\alpha(\textbf{r})=1$ by construct. The KH approximation consistently accounts for both electrostatic and non-polar (associative and steric) effects of solvation in simple and complex liquids, non-electrolyte and electrolyte solutions, compounds and supramolecular solutes in various chemical \cite{Kovalenko:1999:110:10095, Kovalenko:2000:112:10391, Kovalenko:2000:112:10403, Kovalenko:2003:169, Gusarov:2012:33:1478, Kovalenko:2013:85:159, Gusarov:2006:110:6083, Casanova:2007:3:458, Miyata:2008:29:871, Kaminski:2010:114:6082, Malvaldi:2009:113:3536, Kovalenko:2012:8:1508, Kovalenko:2001:349:496, Kovalenko:2002:1:381, Yoshida:2002:106:5042, Omelyan:2003:2:193, Huang:2015:119:5588, Shapovalov:2000:320:186, Stoyanov:2011:203, Fafard:2013:117:18556, Huang:2014:118:23821}, soft matter \cite{Kovalenko:2012:8:1508}, synthetic organic supramolecular \cite{Kovalenko:2013:85:159, Moralez:2005:127:8307, Johnson:2007:129:5735, Yamazaki:2010:11:361}, biopolymeric \cite{Silveira:2013:135:19048, Stoyanov:2014:29:144, Silveira:2015:6:206}, and biomolecular \cite{Kovalenko:2013:85:159, Luchko:2010:6:607, Omelyan:2013:39:25, Omelyan:2013:139:244106, Omelyan:2015:11:1875, Yoshida:2006:128:12042, Yoshida:2009:113:873, Imai:2009:131:12430, Phongphanphanee:2008:130:1540, Phongphanphanee:2010:132:9782, Maruyama:2011:3:290, Yonetani:2008:128:186102, Maruyama:2010:114:6464, Harano:2001:114:9506, Imai:2001:59:512, Imai:2002:112:9469, Imai:2007:16:1927, Kovalenko:2015:22:575, Blinov:2010:98:282, Yamazaki:2008:95:4540, Blinov:2011:37:718, Genheden:2010:114:8505, Kovalenko:2011:164:101, Nikolic:2012:8:3356, Imai:2011:115:8288, Huang:2015:55:317, Stumpe:2011:115:319} systems.

The 3D-KH closure underestimates the height of strong associative peaks of the 3D site distribution functions due to the MSA linearization applied to them \cite{Kovalenko:1999:110:10095, Perkyns:2010:132:064106}. However, it somewhat widens the peaks and so 3D-RISM-KH quite accurately reproduces the coordination numbers of the solvation structure in different systems, including micromicelles in water-alcohol solutions \cite{Kovalenko:2013:85:159, Yoshida:2002:106:5042, Omelyan:2003:2:193, Huang:2015:119:5588}, solvation shells of metal-water \cite{Kovalenko:1999:110:10095, Kovalenko:2003:169}, metal oxide-water \cite{Shapovalov:2000:320:186} and mixed organic solvent-clay \cite{Fafard:2013:117:18556, Huang:2014:118:23821} interfaces, and structural water solvent localized in biomolecular confinement \cite{Yoshida:2009:113:873, Huang:2015:55:317, Stumpe:2011:115:319}. For instance, the coordination numbers of water strongly bound to the MgO surface are calculated from the 3D-RISM-KH theory with a 90\% accuracy and the peak positions within a $0.5~\AA$ deviation, compared to molecular dynamics (MD) simulations \cite{Shapovalov:2000:320:186}. The 3D solvation map $g_\gamma(\textbf{r})$ of function-related structural water in the GroEL chaperon complex (shown to be strongly correlated to the rate of protein folding inside the chaperon cavity) obtained in an expensive MD simulation with explicit solvent involving ~1 million atoms is reproduced from the 3D-RISM-KH theory in a relatively short calculation on a workstation with an accuracy of over 90\% correlation for the 3D density map and about 98\% correlation for the 3D density maxima \cite{Stumpe:2011:115:319}.

Note that different approximate 3D bridge functions, or bridge corrections, to the 3D-HNC approximation can be constructed, such as the 3D-HNCB closure which reproduced MD simulation results for the height of the 3D distribution peaks of water solvent around the bovine pancreatic trypsin inhibitor (BPTI) protein \cite{Perkyns:2010:132:064106}. A related promising approach is the partial series expansions of order n (PSE-n) of the HNC closure \cite{Kast:2008:129:236101}. The PSE-n closures interpolate between the KH and HNC approximations and, therefore, combine numerical stability with enhanced accuracy, as demonstrated for aqueous solutions of alkali halide ions \cite{Kast:2008:129:236101, Joung:2013:138:044103}. However, a particular appeal of the 3D-KH closure is that it provides an adequate accuracy and the existence of solutions if expected from physical considerations for a wide class of solution systems ``from the wild'', including various chemical and biomolecular solutes in solution with different solvents, co-solvents, and electrolytes \cite{Kovalenko:1999:110:10095, Kovalenko:2000:112:10391, Kovalenko:2000:112:10403, Kovalenko:2003:169, Gusarov:2012:33:1478, Kovalenko:2013:85:159, Gusarov:2006:110:6083, Casanova:2007:3:458, Miyata:2008:29:871, Kaminski:2010:114:6082, Malvaldi:2009:113:3536, Kovalenko:2012:8:1508, Kovalenko:2001:349:496, Kovalenko:2002:1:381, Yoshida:2002:106:5042, Omelyan:2003:2:193, Huang:2015:119:5588, Shapovalov:2000:320:186, Stoyanov:2011:203, Fafard:2013:117:18556, Huang:2014:118:23821, Moralez:2005:127:8307, Johnson:2007:129:5735, Yamazaki:2010:11:361, Silveira:2013:135:19048, Stoyanov:2014:29:144, Silveira:2015:6:206, Luchko:2010:6:607, Omelyan:2013:39:25, Omelyan:2013:139:244106, Omelyan:2015:11:1875, Yoshida:2006:128:12042, Yoshida:2009:113:873, Imai:2009:131:12430, Phongphanphanee:2008:130:1540, Phongphanphanee:2010:132:9782, Maruyama:2011:3:290, Yonetani:2008:128:186102, Maruyama:2010:114:6464, Harano:2001:114:9506, Imai:2001:59:512, Imai:2002:112:9469, Imai:2007:16:1927, Kovalenko:2015:22:575, Blinov:2010:98:282, Yamazaki:2008:95:4540, Blinov:2011:37:718, Genheden:2010:114:8505, Kovalenko:2011:164:101, Nikolic:2012:8:3356, Imai:2011:115:8288, Huang:2015:55:317, Stumpe:2011:115:319}, and moreover, with ligand molecules or fragments considered as part of solvent for efficient 3D mapping of binding affinities in such problems as molecular recognition and fragment based drug design \cite{Yoshida:2009:113:873, Nikolic:2012:8:3356, Imai:2011:115:8288}. Construction of bridge functions for such systems with different solvent components other than water constitutes a significant challenge not addressed so far, and the 3D-KH closure offers a reliable choice for a given system.

\subsection{Dielectrically consistent RISM theory for site-site correlations of solvent}
\label{sec:DRISM}

The radially dependent site-site susceptibility of bulk solvent breaks up into the intra- and intermolecular terms,
\begin{equation}
  \chi_{\alpha\gamma}(r) = \omega_{\alpha\gamma}(r) + \rho_\alpha h_{\alpha\gamma}(r) \, .
\label{eq:chi}
\end{equation}
The intramolecular correlation function $\omega_{\alpha\gamma}(r)$ normalized as $4\pi \int r^2 \textrm{d}r \, \omega_{\alpha\gamma}(r) = 1$ represents the geometry of solvent molecules. [Note that $\omega_{\alpha\gamma}(r)=0$ for sites $\alpha$ and $\gamma$ on different species.] For rigid species with site separations $l_{\alpha\gamma}$, it is represented in the direct space in terms of $\delta$-functions as $\omega_{\alpha\gamma}(r) = \delta_{\alpha\gamma}(r-l_{\alpha\gamma}) / ( 4\pi l_{\alpha\gamma}^2 )$ and for numerical calculations it is specified in reciprocal $k$-space as
\begin{equation}
  \omega_{\alpha\gamma}(k) = j_0\left(kl_{\alpha\gamma}\right) \, ,
\label{eq:omega}
\end{equation}
where $j_0(x)$ is the zeroth-order spherical Bessel function. The intermolecular term in \eqref{eq:chi} to be input into \eqref{eq:3D-RISM} is given by the site-site radial total correlation function $h_{\alpha\gamma}(r)$ between all sites $\alpha$ and $\gamma$ on all sorts of molecules in bulk solvent. The latter are obtained in advance to the 3D-RISM-KH calculation from the dielectrically consistent RISM theory \cite{Perkyns:1992:190:626, Perkyns:1992:97:7656} coupled with the KH closure (DRISM-KH approach) \cite{Kovalenko:2003:169, Kovalenko:2013:85:159} which can be applied to a bulk solution of a given composition, including polar solvent, co-solvent, electrolyte, and ligands at a given concentration. The DRISM integral equation reads \cite{Perkyns:1992:190:626, Perkyns:1992:97:7656}
\begin{subequations}
\begin{equation}
\label{eq:DRISM}
  \tilde{h}_{\alpha\gamma}(r) = \tilde{\omega}_{\alpha\mu}(r) \ast c_{\mu\nu}(r)
                                \ast \left[ \tilde{\omega}_{\nu\gamma}(r) + \rho_\nu \tilde{h}_{\nu\gamma}(r) \right],
\end{equation}
where $c_{\alpha\gamma}(r)$ is the site-site direct correlation function of pure bulk solvent, and both the intramolecular correlation function $\tilde{\omega}_{\alpha\gamma}(r)$ and the total correlation function $\tilde{h}_{\alpha\gamma}(r)$ are renormalized due to a dielectric bridge correction in a particular analytical form \cite{Perkyns:1992:190:626, Perkyns:1992:97:7656} that ensures the given phenomenological value as well as consistency of the dielectric constant determined through the three different routes related to the solvent-solvent, solvent-ion, and ion-ion effective interactions in electrolyte solution,
\begin{align}
  \tilde{\omega}_{\alpha\gamma}(r) & = \omega_{\alpha\gamma}(r) + \rho_\alpha \zeta_{\alpha\gamma}(r) \, , \\
       \tilde{h}_{\alpha\gamma}(r) & = h_{\alpha\gamma}(r) - \zeta_{\alpha\gamma}(r) \, .
\end{align}
\end{subequations}
The renormalized dielectric correction enforcing the given phenomenological value of the dielectric constant and the proper orientational behavior and consistency of the dielectric response in the electrolyte solution is obtained in the analytical form specified in the reciprocal $k$-space as follows \cite{Perkyns:1992:190:626, Perkyns:1992:97:7656}, \begin{equation}
  \zeta_{\alpha\gamma}(k) = j_0(k x_\alpha) j_0(k y_\alpha) j_1(k z_\alpha) h_{\textrm{c}}(k) j_0(k x_\gamma) j_0(k y_\gamma) j_1(k z_\gamma) \, ,
\label{eq:DielCorrZeta}
\end{equation}
where $j_0(x)$ and $j_1(x)$ are the zeroth- and first-order spherical Bessel functions, $\textbf{r}_\alpha=(x_\alpha , y_\alpha , z_\alpha)$ are the Cartesian coordinates of partial charge $q_\alpha$ of site $\alpha$ on species $s$ with respect to its molecular origin; both sites $\alpha$ and $\gamma$ are on the same species $s$, and its dipole moment $\textbf{ d}_s = \sum_{\alpha\in s} q_\alpha \textbf{r}_\alpha$ is oriented along the $z$-axis: $\textbf{d}_s=(0,0,d_s)$. Note that the renormalized dielectric correction \eqref{eq:DielCorrZeta} is nonzero only for polar solvent species of electrolyte solution which possess a dipole moment and thus are responsible for the dielectric response in the DRISM approach. The envelope function $h_{\textrm{c}}(k)$ has the value at $k=0$ determining the dielectric constant of the solution and is assumed in a smooth non-oscillatory form quickly falling off at wavevectors k larger than the inverse characteristic size $l$ of liquid molecules \cite{Perkyns:1992:190:626, Perkyns:1992:97:7656},
\begin{equation}
  h_{\textrm{c}}(k) = A \exp \left( - l^2 k^2 / 4 \right) \, ,
\label{eq:DielCorrHc}
\end{equation}
where $A=(y^{-1}\epsilon-3)\rho_{\textrm{polar}}^{-1}$ is the amplitude related to the dielectric constant $\epsilon$ of the electrolyte solution which is a phenomenological parameter specified at input. The form (\ref{eq:DielCorrZeta}) has also been extended to solvent mixtures \cite{Kvamme:1993:16:743}, where the total number density of polar species $\rho_{\textrm{polar}} = \sum_{s\in{\textrm{polar}}} \rho_s$ and the dielectric susceptibility of the mixture is given in the general case by
\begin{equation}
  y = \frac{4\pi}{9 k_{\textrm{B}}T}\sum\limits_{a\in{\textrm{polar}}} \rho_a d_a^2\,.
\label{eq:DielCorrSusc}
\end{equation}
The parameter $l$ in the envelope function (\ref{eq:DielCorrHc}) specifies the characteristic separation from a molecule in solution below which the dielectric correction (\ref{eq:DielCorrZeta}) is switched off so as not to distort the short-range solvation structure. It can be chosen to about $l=1$~{\AA} for water solvent; however, in solvent of larger molecules or in the presence of such co-solvent, it should be increased  so as to avoid ``ghost'' associative peaks appearing in the radial distributions if the dielectric correction (\ref{eq:DielCorrZeta}) interferes with the intramolecular structure of the large solvent species. For example, for octanol solvent, it should be set to about $l=10$~{\AA}.

The DRISM integral equation (\ref{eq:DRISM}) is complemented with the KH closure \cite{Kovalenko:2003:169, Kovalenko:2013:85:159} which in the site-site radial version reads
\begin{equation}
  g_{\alpha\gamma}(r) =
    \left\{
    \begin{array}{lcc}
      \exp\bigl[ - u_{\alpha\gamma}(r) / (k_{\textrm{B}}T) + h_{\alpha\gamma}(r) - c_{\alpha\gamma}(r) \bigr]
      & \textrm{for} & g_{\alpha\gamma}(r) \leqslant 1,
    \\ [6pt]
      1 - u_{\alpha\gamma}(r) / (k_{\textrm{B}}T) + h_{\alpha\gamma}(r) - c_{\alpha\gamma}(r)
      & \textrm{for} & g_{\alpha\gamma}(r) > 1,
   \end{array} \right.
\label{eq:KH}
\end{equation}
where the site-site interaction potential $u_{\alpha\gamma}(r)$ (all-atom force field) is typically modelled  with Coulomb and Lennard-Jones terms. The DRISM-KH equations (\ref{eq:DRISM}) and (\ref{eq:KH}) keep the same dielectrically consistent asymptotics (\ref{eq:DielCorrZeta}) as the original DRISM-HNC theory \cite{Perkyns:1992:190:626, Perkyns:1992:97:7656}, but extend the description to solutions with strong associative species in a wide range of compositions and thermodynamic conditions not amenable to the HNC closure.
(The latter overestimates the associative attraction and density inhomogeneities and usually diverges for such systems.)

\subsection{Analytical expressions for solvation thermodynamics}
\label{sec:Thermodynamics}

Much as the HNC approximation, the 3D-KH closure \eqref{eq:3D-KH} to the 3D-RISM integral equation \eqref{eq:3D-RISM} has an exact differential of the solvation free energy, and thus allows one to analytically perform Kirkwood's thermodynamic integration gradually switching on the solute-solvent interaction. This gives the solvation free energy of the solute molecule (or supramolecule) in multicomponent solvent in a closed analytical form in terms of the 3D site total and direct correlation functions $h_\gamma(\textbf{r})$ and $c_\gamma(\textbf{r})$ \cite{Kovalenko:1999:110:10095, Kovalenko:2003:169, Kovalenko:2013:85:159}:
\begin{subequations}
\label{eq:SFE-KH}
\begin{align}
\label{eq:SFE-KH-int}
 \Delta\mu
 & = \sum_\gamma \displaystyle\int_V \textrm{d}\textbf{r} \, \Phi_\gamma^{\textrm{KH}}(\textbf{r}) \, ,
  \\
 \Phi_\gamma^{\textrm{KH}}(\textbf{r})
&  = \rho_\gamma k_{\textrm{B}}T
    \left[ \frac12 \big(h_\gamma(\textbf{r})\big)^2 \Theta\big(-h_\gamma(\textbf{r})\big)
    - \frac12 h_\gamma(\textbf{r}) c_\gamma(\textbf{r}) - c_\gamma(\textbf{r}) \right] \, ,
\label{eq:SFED-KH}
\end{align}
\end{subequations}
where the sum goes over all the sites of all solvent species, and $\Theta(x)$ is the Heaviside step function. This offers several substantial advantages over molecular simulations using free energy perturbation techniques with multiple productive runs to perform the thermodynamic integration or solute mutation, which is extremely time consuming \cite{Chipot:2007}. Once the 3D-RISM-KH equations \eqref{eq:3D-RISM}, \eqref{eq:3D-KH} are converged for the correlation functions $c_\alpha(\textbf{r})$ and $h_\alpha(\textbf{r})$, all the solvation thermodynamics is readily available from the expression \eqref{eq:SFE-KH} and analysis of its components. Note that accurate calculation of the solvation free energy from the functional \eqref{eq:SFE-KH} requires proper analytical account of electrostatic asymptotics in all the correlation functions, both 1D and 3D, in the integral equations \eqref{eq:3D-RISM}--\eqref{eq:KH} \cite{Kovalenko:2000:112:10391,Kovalenko:2000:112:10403, Kovalenko:2003:169, Gusarov:2012:33:1478, Kovalenko:2013:85:159}.

The integrand $\Phi_\gamma^{\textrm{KH}}(\textbf{r})$ in \eqref{eq:SFE-KH-int} is interpreted as the solvation free energy density (3D-SFED) coming from interaction site $\gamma$ of solvent molecules around the solute. The solvation free energy of the solute macromolecule $\Delta\mu$ is obtained by summation of the 3D-SFED partial contributions from all solvent species and spatial integration over the whole space. Note that the integration volume $V$ in the form \eqref{eq:SFE-KH} comprises both the solvation shells and the solute-solvent molecular repulsive cores, since the direct correlation functions $c_\alpha(\textbf{r})$ inside the latter are related to the free energy of creation of a cavity to accommodate the solute excluded volume. In this way, $\Phi_\gamma^{\textrm{KH}}(\textbf{r})$ characterizes the intensity of effective solvation forces in different 3D spatial regions of the solvation shells and indicates where they contribute the most/least to the entire solvation free energy.

The solvation free energy in the form \eqref{eq:SFE-KH} can be split up into components from each solvent species by grouping the terms for the corresponding sites. Further, integrating the 3D-SFED values $\Phi_\gamma^{\textrm{KH}}(\textbf{ r})$ over space regions geometrically attributed to the constituting chemical groups of the solute by a Voronoi decomposition (including the repulsive core region) resolves partial contributions of functional groups of the solute molecule (or supramolecule) to the solvation free energy, which is called spatial decomposition analysis \cite{Yamazaki:2009:5:1723, Yamazaki:2011:115:310}.

With the solvation free energy obtained in the form \eqref{eq:SFE-KH}, the potential of mean force $w(\textbf{1},\textbf{n})$ between $i=1,n$ solute molecules (or equally, supramolecule parts or nanoparticles) in solution at given distances and orientations in an arrangement $(\textbf{1},\textbf{n})$ is calculated directly by definition as their interaction potential $u(\textbf{1},\textbf{n})$ plus the change in the solvation free energy from the sum of $\Delta\mu(\textbf{i})$ of separate particles to non-additive $\Delta\mu(\textbf{1},\textbf{n})$ of the particles brought together \cite{Kovalenko:2000:112:10391, Kovalenko:2000:112:10403, Kovalenko:2003:169, Gusarov:2012:33:1478, Kovalenko:2013:85:159},
\begin{equation}
  w(\textbf{1},\textbf{n}) = u(\textbf{1},\textbf{n}) + \Delta\mu(\textbf{1},\textbf{n}) - \sum\limits_{i=1}\Delta\mu(\textbf{i}) \, .
\label{eq:PMF-dir}
\end{equation}
Similarly to the solvation free energy \eqref{eq:SFE-KH}, the PMF \eqref{eq:PMF-dir} can be split up by spatial decomposition analysis into contributions of solute chemical functionalities and into terms mediated by each solvent species. This route provides an efficient and accurate statistically representative way to calculate effective interactions of various chemical species \cite{Kovalenko:2000:112:10391, Kovalenko:2000:112:10403, Kovalenko:2003:169, Gusarov:2012:33:1478, Kovalenko:2013:85:159, Stoyanov:2011:203, Fafard:2013:117:18556}, nanomaterials \cite{Moralez:2005:127:8307, Johnson:2007:129:5735, Yamazaki:2010:11:361, Silveira:2013:135:19048, Stoyanov:2014:29:144, Silveira:2015:6:206}, and biomolecular nanosystems\cite{Blinov:2010:98:282, Yamazaki:2008:95:4540, Blinov:2011:37:718, Genheden:2010:114:8505, Kovalenko:2011:164:101}.

Furthermore, PMFs between a solute molecule / supramolecule / nanoparticle and solvent molecules averaged over molecular orientations centered at solvent interaction sites can be conveniently obtained in the 3D site representation from the definition of the 3D site density distribution functions,
\begin{equation}
  w_\gamma(\textbf{r}) = - k_{\textrm{B}}T \ln\left[ g_\gamma(\textbf{r}) \right] \, .
\label{eq:PMF-g}
\end{equation}
The solute-solvent PMFs so obtained in a single 3D-RISM-KH calculation provide 3D spatial maps of site binding affinity of solvent species to the solute molecule (positive in attractive wells and negative in repulsive barriers of PMFs). This constitutes a new approach toward molecular recognition and computational fragment-based drug design \cite{Yoshida:2009:113:873}. With fragmental decomposition of flexible ligands treated as distinct species in solvent mixture of arbitrary complexity, the computed density functions and the corresponding PMFs \eqref{eq:PMF-g} of solvent, ions, and ligand fragments around a biomolecule obtained from the 3D-RISM-KH theory on discrete 3D spatial grids uniquely define the scoring function which can be interfaced to a docking protocol \cite{Yoshida:2009:113:873, Imai:2009:131:12430, Kovalenko:2011:164:101, Nikolic:2012:8:3356}, e.g., in the 3D-RISM-Dock approach implemented in the AutoDock suite for automated ranking of docked conformations \cite{Nikolic:2012:8:3356}. As a validation, the 3D-RISM-Dock tool predicted the location and residency times of the modes of binding of a flexible thiamine molecule to the prion protein at near-physiological conditions in an excellent agreement with experiment \cite{Kovalenko:2011:164:101, Nikolic:2012:8:3356}. The capabilities of 3D maps of ligand binding affinity calculated from \eqref{eq:PMF-g} include the prediction of functions in biomolecular systems, such as drug efflux pathway in multidrug transporter AcrB \cite{Imai:2011:115:8288}.

Further, the solvation free energy \eqref{eq:SFE-KH} can be split up into the energetic and entropic contributions \cite{HansenMcDonald:2006, Yu:1988:89:2366, Yu:1990:92:5020},
\begin{equation}
  \Delta\mu = \Delta\varepsilon^{\textrm{uv}} + \Delta\varepsilon^{\textrm{vv}} - T\Delta s_V \, ,
\label{eq:SFE-Energy-Entropy}
\end{equation}
by calculating the solvation entropy at constant volume as
\begin{equation}
  \Delta s_V = - \frac{1}{T} \left(\frac{\partial\Delta\mu}{\partial T}\right)_V
\label{eq:SFE-Entropy}
\end{equation}
and the internal energy of the solute (``u'') --- solvent (``v'') interaction as
\begin{equation}
  \Delta\varepsilon^{\textrm{uv}} = - k_{\textrm{B}}T \sum\limits_\gamma \int\textrm{d}\textbf{r} \, g_\gamma(\textbf{r}) \, u_\gamma(\textbf{r}) \, ,
\label{eq:SFE-Energy-uv}
\end{equation}
where the remaining term $\Delta\varepsilon^{\textrm{vv}}$ gives the energy of solvent reorganization around the solute molecule. In the same way, the PMF \eqref{eq:PMF-dir} or \eqref{eq:PMF-g} can be decomposed into energetic and entropic terms.

Further to the solvation free energy, the 3D-RISM-KH theory also provides analytical expressions for other thermodynamic functions, including the partial molar volume (PMV) of the solute molecule obtained in the Kirkwood-Buff theory in terms of the 3D direct correlation functions from the 3D-RISM integral equation \cite{Harano:2001:114:9506, Imai:2001:59:512, Imai:2002:112:9469, Imai:2007:16:1927, Kovalenko:2015:22:575},
\begin{equation}
  \bar{V} = k_{\textrm{B}}T \chi_T \left( 1 - \sum\limits_\gamma \rho_\gamma
                                             \int\textrm{d}\textbf{r} \, c_\gamma(\textbf{r}) \right) \, ,
\label{eq:PMV}
\end{equation}
where $\chi_T$ is the isothermal compressibility of bulk solvent which is analytically obtained in the so-called compressibility route from the RISM integral equation as \cite{Yoshida:2002:106:5042}
\begin{equation}
  k_{\textrm{B}}T \chi_T = \rho^{-1} \left( 1 - 4\pi\sum\limits_{\alpha\gamma} \rho_\alpha
                           \int\limits_0^\infty r^2 \textrm{d}r \, c_{\alpha\gamma}(r) \right)^{-1}
\label{eq:compres}
\end{equation}
in terms of the site-site direct correlation functions of bulk solvent $c_{\alpha\gamma}(r)$ calculated along with the solvent susceptibility $\chi_{\alpha\gamma}(r)$ from the DRISM-KH equations, and $\rho=\sum_s \rho_s$ is the total number density of species $s$ in the solution bulk.

As noted above, the solvation free energy expression \eqref{eq:SFE-KH} follows from the functional form of the KH closure \eqref{eq:3D-KH} to the 3D-RISM integral equation \eqref{eq:3D-RISM} by carrying out the thermodynamic integration analytically. In practice, the 3D-RISM-KH theory can be used with a different functional of the solvation free energy. In particular, Chandler and co-workers derived the so-called Gaussian Fluctuation (GF) free energy functional based on the assumption of Gaussian fluctuations for the solvent, which is similar to the HNC functional form but without the $h_\gamma^2$ term \cite{Chandler:1984:81:1975, Chandler:1993:48:2898}. The solvation free energy calculated from the GF functional using site-site correlation functions obtained from the RISM-HNC integral equations were generally in reasonable agreement with experiment and in better agreement than those obtained with the HNC functional \cite{Ichiye:1988:92:5257, Lee:1993:97:10175}. In the context of 3D-RISM theory, the GF functional reads \cite{Luchko:2010:6:607, Genheden:2010:114:8505}
\begin{subequations}
\label{eq:SFE-GF}
\begin{align}
\label{eq:SFE-GF-int}
 \Delta\mu
&  = \sum_\gamma \displaystyle\int_V \textrm{d}\textbf{r} \, \Phi_\gamma^{\textrm{GF}}(\textbf{r}) \, ,
  \\
 \Phi_\gamma^{\textrm{GF}}(\textbf{r})
&  = \rho_\gamma k_{\textrm{B}}T
    \left[ - \frac12 h_\gamma(\textbf{r}) c_\gamma(\textbf{r}) - c_\gamma(\textbf{r}) \right] \, .
\label{eq:SFED-GF}
\end{align}
\end{subequations}
The GF functional using the 3D correlation functions obtained from the 3D-RISM-KH theory provides the hydration free energy in better agreement with experiment in some cases \cite{Genheden:2010:114:8505, Palmer:2010:22:492101}.

Recently, it was found that the 3D-RISM-KH theory supplemented with the solvation free energy correction based on the partial molar volume of the compound (also obtained from the 3D-RISM-KH theory) provides the hydration free energy for a large diverse set of compounds with an accuracy comparable to much more computationally demanding molecular dynamics simulations with explicit solvent \cite{Palmer:2010:22:492101}. The correction to the solvation free energy $\Delta\mu$ calculated from either the KH functional \eqref{eq:SFE-KH} or the GF one \eqref{eq:SFE-GF} is constructed as a linear function of the PMV in turn calculated in the form \eqref{eq:PMV} using the correlation functions from the 3D-RISM-KH theory,
\begin{equation}
  \Delta\mu_{\textrm{UC}} = \Delta\mu + \alpha\left(\rho\bar{V}\right) + \beta \, .
\label{eq:SFE-UC}
\end{equation}
The constants $\alpha$ and $\beta$ in \eqref{eq:SFE-UC} are obtained from the linear regression analysis. This partial molar volume correction technique, also called a ``Universal Correction'' (UC), was parameterized with the GF functional on hydration free energies of a training set of 65 molecules and was tested on a set of 120 molecules \cite{Palmer:2010:22:492101}. The UC to the KH functional has also been parameterized for the hydration free energy on a set of 504 organic molecules \cite{Truchon:2014:10:934}. In the same study, another correction to the KH functional based on the Ng modified free energy functional \cite{Ng:1974:61:2680} was introduced and shown to accurately predict hydration free energies of the same set of 504 compounds \cite{Truchon:2014:10:934}.

With the partial molar volume correction (or the UC term) re-parameterized accordingly, the 3D-RISM-KH theory also provides an excellent agreement with the experimental data for the solvation free energy in non-polar solvent (1-octanol) of a large library of 172 small compounds with diverse functional groups, and so accurately predicts the octanol-water partition coefficients \cite{Huang:2015:119:5588}. The best agreement with the experimental data for octanol-water partition coefficients is obtained with the KH-UC solvation free energy functional.

\subsection{Analytical treatment of electrostatic asymptotics}
\label{sec:Asymptotics}

To properly treat electrostatic forces in electrolyte solution with polar solvent and ionic species in 3D-RISM / DRISM theory requires analytical treatment of the electrostatic asymptotics of the radial site-site total and direct correlation functions of bulk solvent in the DRISM-KH equations \eqref{eq:DRISM}, \eqref{eq:KH}, as well as of the 3D site total and direct correlation functions in the 3D-RISM-KH equations \eqref{eq:3D-RISM}, \eqref{eq:3D-KH}, the solvation free energy functional \eqref{eq:SFE-KH} or \eqref{eq:SFE-GF}, its derivatives \eqref{eq:PMF-dir}--\eqref{eq:SFE-Energy-uv}, and the volumetrics \eqref{eq:PMV}--\eqref{eq:compres} \cite{Kovalenko:2000:112:10391, Kovalenko:2000:112:10403, Kovalenko:2003:169, Gusarov:2012:33:1478, Kovalenko:2013:85:159, Kaminski:2010:114:6082}. The convolution in the 3D-RISM integral equation \eqref{eq:3D-RISM} is calculated by using the 3D fast Fourier transform (3D-FFT) technique. The DRISM-KH equations are discretized on a uniform radial grid of resolution $0.01-0.1~\AA$, and the 3D-RISM-KH equations are discretized on a uniform 3D rectangular grid with resolution $0.2-0.5~\AA$ in a 3D box of size including 2--3 solvation shells around the solute (supra)molecule. (Typical box grid sizes range from 64$\times$64$\times$64 to 256$\times$256$\times$256 nodes.) Analytical forms of the non-periodic electrostatic asymptotics in the direct and reciprocal space are separated out from all the correlation functions before and then added back after the 3D-FFT. Note that although the solvent susceptibility $\chi_{\alpha\gamma}(r)$ has a long-range electrostatic part, no aliasing occurs in the backward 3D-FFT of the short-range part of $h_\gamma(\textbf{k})$ on the 3D box supercell since the short-range part of the convolution product $h_\gamma(\textbf{r})$ contains merely 2--3 oscillations (solvation shells) and vanishes outside the 3D box for physical reasons \cite{Gusarov:2012:33:1478, Kovalenko:2013:85:159, Kaminski:2010:114:6082}. Accordingly, the electrostatic asymptotics terms in the thermodynamic integral \eqref{eq:SFE-KH} are handled analytically and reduced to one-dimensional integrals easy to compute \cite{Kovalenko:2000:112:10391, Kovalenko:2000:112:10403, Kovalenko:2003:169, Gusarov:2012:33:1478, Kovalenko:2013:85:159, Kaminski:2010:114:6082}.

\subsection{Accelerated numerical solution}
\label{sec:MDIIS}

The 3D-RISM-KH integral equations \eqref{eq:3D-RISM}, \eqref{eq:3D-KH} are converged typically to a root mean square accuracy of $10^{-4}-10^{-8}$ and the DRISM-KH equations \eqref{eq:DRISM}, \eqref{eq:KH} to an accuracy of $10^{-6}-10^{-12}$ by using the modified direct inversion in the iterative subspace (MDIIS) accelerated numerical solver \cite{Kovalenko:1999:20:928, Kovalenko:2000:112:10391, Kovalenko:2000:112:10403, Kovalenko:2003:169, Gusarov:2012:33:1478}. MDIIS is an iterative procedure accelerating the convergence of integral equations of liquid state theory by optimizing each iterative guess in a Krylov subspace of typically last 10--20 successive iterations and then making the next iterative guess by mixing the optimized solution with the approximated optimized residual \cite{Kovalenko:1999:20:928,Kovalenko:2000:112:10391, Kovalenko:2003:169, Gusarov:2012:33:1478}. It is closely related to Pulay's DIIS approach for quantum chemistry equations \cite{Pulay:1980:73:393}. and other similar algorithms like the generalized minimal residual (GMRes) solver \cite{Saad:1986:7:856}. The MDIIS solver combines the simplicity and relatively small memory usage of an iterative approach with the efficiency of a direct method. Compared to damped (Picard) iterations, MDIIS provides substantial acceleration with quasiquadratic convergence practically throughout the whole range of root mean square residual, and is robust and stable. Of particular importance is that MDIIS ensures convergence (provided a solution exists) for complex charged systems with strong associative and steric effects, which is usually not achievable with Picard iterations and constitutes a challenging task in the case of 3D integral equations on large 3D grids. Converging the 3D-RISM-KH equations takes minutes to hours on a HPC workstation, depending on the 3D grid size, whereas the DRISM-KH equations converge in seconds for water but much slower, up to hours, for multi-component solvent mixtures with complex species.

\section{Examples of 3D-RISM-KH calculations for the solvation structure and thermodynamics of solution systems and nanoparticles}
\label{sec:Examples}

Sketched below are the capabilities of the 3D-RISM-KH molecular theory of solvation in predictive modelling of some representative molecules and nanosystems in solution: structure and potentials of mean force in aqueous electrolyte solutions, activity coefficients of simple and molecular ions in electrolyte solutions, structural transitions in mixtures of water and amphiphile co-solvent, compound partitioning between water and octanol solvents, and effective nanoscale forces between cellulose nanoparticles in hemicellulose hydrogel maintaining the resilient structure of plant cell walls. The discussion below concerns the main features, capabilities and conclusions, whereas the force fields used and different setup details can be found in the corresponding original work cited here.

\subsection{Simple ions in aqueous electrolyte solution}
\label{sec:AqueousElectrolyte}

\begin{figure}[!htb]
\centering
\includegraphics[width=1.0\textwidth]{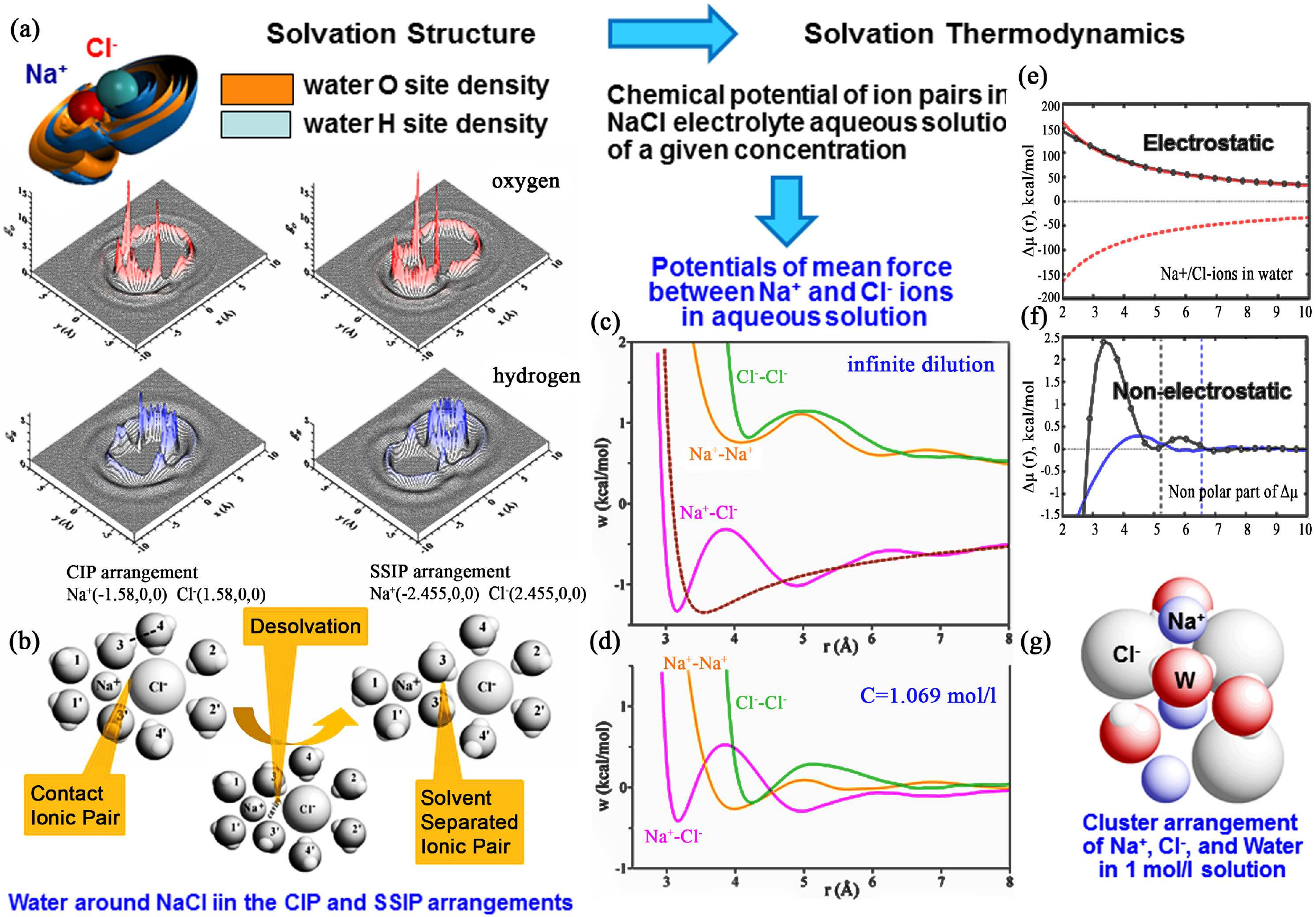}
\caption{
\label{fig:1}
(Color online) Solvation structure of the contact ion pair (CIP) and solvent separated ion pair (SSIP) of the Na$^+$ and Cl$^-$ ion pair in aqueous solution at infinite dilution, obtained from 3D-RISM theory \cite{Kovalenko:2000:112:10391,Kovalenko:2000:112:10403}. Part (a): Section of the 3D distribution functions of water oxygen (O) and hydrogen (H) in the plane passing through the ion-ion axis. Part (b): Visualization of the water solvent arrangements around the CIP and SSIP, as well as the ions separated by a gap corresponding to the barrier at their potential of mean force. Part (c): Potentials of mean force calculated using \eqref{eq:PMF-dir} for all the pairs of Na$^+$ and Cl$^-$ ions in aqueous electrolyte solution at infinite dilution and at concentration 1.06 mol/l. Part (d): Shown for comparison is also the potential of mean force of the Na$^+$--Cl$^-$ ion pair in the structureless dielectric continuum (primitive model) of water solvent. Decomposition of the potential of mean force of the ion pairs at infinite dilution into: (d) electrostatic term for the Na$^+$--Na$^+$, Cl$^-$--Cl$^-$, and Na$^+$--Cl$^-$ ion pairs (two upper solid curves and lower dashed curve, respectively);  (e) non-electrostatic (non-polar) term for the Na$^+$--Cl$^-$ ion pair at full ion charges and with ion charges switched off (solid curves with and without symbols, respectively). Part (g): Visualization of a cluster hydrated ions, based on the peaks positions on the 3D maps of density distributions: Cl$^-$ ions (shown in gray) in a CIP arrangement stabilized due to bridging by associated Na$^+$ ions (in blue) as well as by hydrogen bonded water molecules (water O in red, water H in white) in aqueous electrolyte solution of concentration 1M.
}
\end{figure}

One of the basic cases of electrolyte solution systems is aqueous solution of sodium chloride. Figure~\ref{fig:1} presents the 3D-RISM results for the solvation structure and potential of mean force for all the pairs of Na$^+$ and Cl$^-$ ions in aqueous electrolyte solution at infinite dilution and at a relatively high concentration of 1.069 mol/l \cite{Kovalenko:2000:112:10391, Kovalenko:2000:112:10403}. This generic system presents an illustrative test for a solvation theory to reproduce most of the essential features of a variety of chemical and biomolecular effects in solution, and the 3D-RISM theory succeeds in that.

The 3D site distributions of the water oxygen (O) and hydrogen (H) site distributions around the pair of the Na$^+$ and Cl$^-$ ions in aqueous solution at infinite dilution are shown in figure~\ref{fig:1}~(a). Both the O and H distributions form high crowns of the first solvation shell around the ions, with the high O and lower H peaks near the contacts of the crowns corresponding to water molecules bridging the ions. This structure is then followed by the shallow second solvation shells. The corresponding potential of mean force (PMF) calculated using equation~\eqref{eq:PMF-dir} is shown in figure~\ref{fig:1}~(c). The H distribution has two crowns around the Cl$^-$ ion; their separation from the ion shows that the inner H crown gives the water hydrogens in contact with the Cl$^-$ ion while the outer H crown corresponds to the other water hydrogens looking outwards but at the angle determined by the tetrahedral hydrogen bonding structure of the water. On the other hand, there is a single H crown around the Na$^+$ ion and the separations of the O and H crowns from Na$^+$ show that both the water hydrogens are looking outward Na$^+$, tilted at the same angle. These are typical arrangements of water around a cation oriented like a dipole and water bonded with one hydrogen to an anion. The arrangements are visualized in the cartoons in figure~\ref{fig:1}~(b) for the 3D water site distributions around both the contact ion pair (CIP) and solvent separated ion pair (SSIP) of the Na$^+$ and Cl$^-$ ions. Water molecules in contact with both the cation and anion form the dipole like association with the former and the hydrogen bonding with the latter, thus creating a water bridge of strongly associated water molecules located in a ring between the ions. This bridge strongly deepens the solvation contribution to the PMF at the CIP arrangement; interplaying with the ion-ion core repulsion, this results in a significant shift of the first minimum on the PMF to a shorter ion-ion separation, compared to the primitive solvation model just uniformly reducing the Coulomb attraction between the ions by the water dielectric constant $\epsilon=80$ [dashed line in figure~\ref{fig:1}~(c)]. At the SSIP arrangement, the water bridge strengthens the association between the ions and results in the second minimum on the ion-ion PMF [figure~\ref{fig:1}~(d)]. Oscillations diminish with distance and the PMF goes to the limit of the pure dielectrically screened electrostatic potential (dashed line). At an intermediate separation between the ions, a desolvated gap forms due to the steric effect of expulsion of solvent molecules by the repulsive cores of the ions [figure~\ref{fig:1}~(b)]. The work against the solvent environment to expel the solvent and create the desolvation gap results in a barrier between the PMF first and second minima corresponding to the CIP and SSIP arrangements [figure~\ref{fig:1}~(c)].

The decomposition of the PMFs for the pairs of Na$^+$ and Cl$^-$ ions is also shown in figure~\ref{fig:1}~(e), (f). The non-electrostatic (or non-polar) part of the PMFs can be defined either as a difference of the full PMF and its Born electrostatic part (solid curve without symbols in the middle right-hand part for the Na$^+$--Cl$^-$ ion pair) or as a PMF obtained from the 3D-RISM calculations for the ion pair solute with the same Lennard-Jones short-range potentials but with the charges of the ions entirely switched off while keeping the charges of water molecules (solid curve with symbols in the middle right-hand part). The difference between these two curves illustrates a strong coupling and interplay of the electrostatic and non-electrostatic forces in the solvation structure. Note that in an aqueous solution or other highly polar solvent, the electrostatic part of the mean solvation forces responsible for dielectric screening always opposes and almost cancels out the Coulomb interaction in the ionic pair and so the resulting PMF comes out effectively scaled down by the high dielectric constant of the polar solvent.

The PMFs in the Na$^+$--Na$^+$ and Cl$^-$--Cl$^-$ pairs of like ions have the same features of the first and second minima and the barrier between them due to the interplay of the associative forces and molecular structure of the ions and solvent molecules. However, at infinite dilution, the strength of the solvent bridges is not sufficient to overcome the electrostatic repulsion and to stabilize the like ion pairs, and both the first and second minima are local [figure~\ref{fig:1}~(c)]. (Note that this can be very different for large molecular ions with weaker electrostatic attraction at the separation determined by their size.) The picture changes for the electrolyte solution at a high concentration of 1 mol/l; numerous salt bridges form in addition to water hydrogen bridges, and the like ion pairs get stabilized in both CIP and SSIP arrangements [figure~\ref{fig:1}~(d)]. For example, the Cl$^-$--Cl$^-$ ion pair in the aqueous solution at this concentration is bridged by several Na$^+$ ions and water molecules, forming a cluster depicted in figure~\ref{fig:1}~(g). The structure of such a cluster follows from the analysis of the 3D-RISM results for the 3D distribution functions of solution species and the corresponding coordination numbers of Na$^+$, Cl$^-$, and water around each ion pair (Na$^+$--Cl$^-$, Cl$^-$--Cl$^-$, and Cl$^-$--Cl$^-$).

\subsection{Activities of ions in electrolyte solution}
\label{sec:Activities}

\begin{figure}[!htb]
\centering
\includegraphics[width=0.5\textwidth]{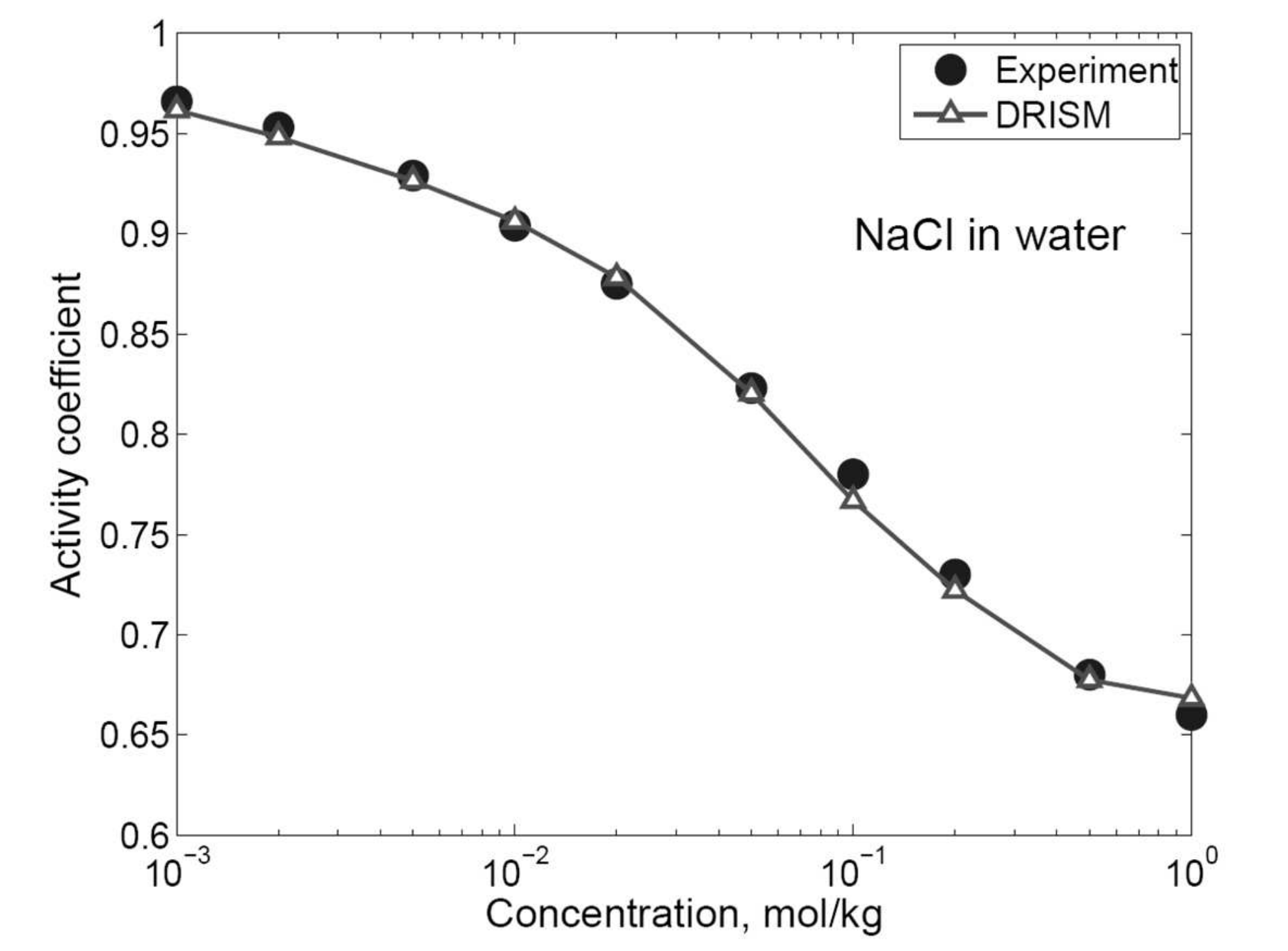}
\caption{
\label{fig:2}
Activity coefficients of ions in ambient aqueous solution of sodium chloride at concentration from low to high against experimental data.
}
\end{figure}

The predictive capability of the 3D-RISM-KH theory (DRISM-KH for simple ions as solutes) in reproducing the solvation thermochemistry of electrolyte solutions in a wide range of concentrations has been validated and shown to be in good agreement with experiment, in particular, for aqueous electrolyte solutions of molecular as well as simple ions in ambient conditions at concentrations from infinite dilution to high ionic strength \cite{Schmeer:2010:12:2407, Joung:2013:138:044103}. In particular, the theory yields the activity coefficient of the ion $\gamma_{\textrm{i}}$ in terms of its solvation free energy $\Delta\mu(c)$ in solution with solvent density $\rho_{\textrm{s}}(c)$ at electrolyte concentration $c$ with respect to those at infinite dilution $c=0$ \cite{Schmeer:2010:12:2407, Joung:2013:138:044103},
\begin{equation}
  \gamma_{\textrm{i}} = \frac{\rho_{\textrm{s}}(c)}{\rho_{\textrm{s}}(c=0)}
                   \exp\left[ \frac{ \Delta\mu(c) - \Delta\mu(c=0) }{ k_{\textrm{B}}T } \right] \, .
\label{eq:Activity}
\end{equation}
For illustration, figure~\ref{fig:2} presents a comparison of the activity coefficient in ambient aqueous solution of sodium chloride at concentrations up to 1 mol/kg calculated from the DRISM-KH theory against experimental data. The calculated activity coefficient is in a good agreement with experiment in the whole range of concentrations.

\subsection{Structure of water-alcohol mixtures}
\label{sec:Water-Alcohol}

\begin{figure}[!htb]
\centering
\includegraphics[width=1.0\textwidth]{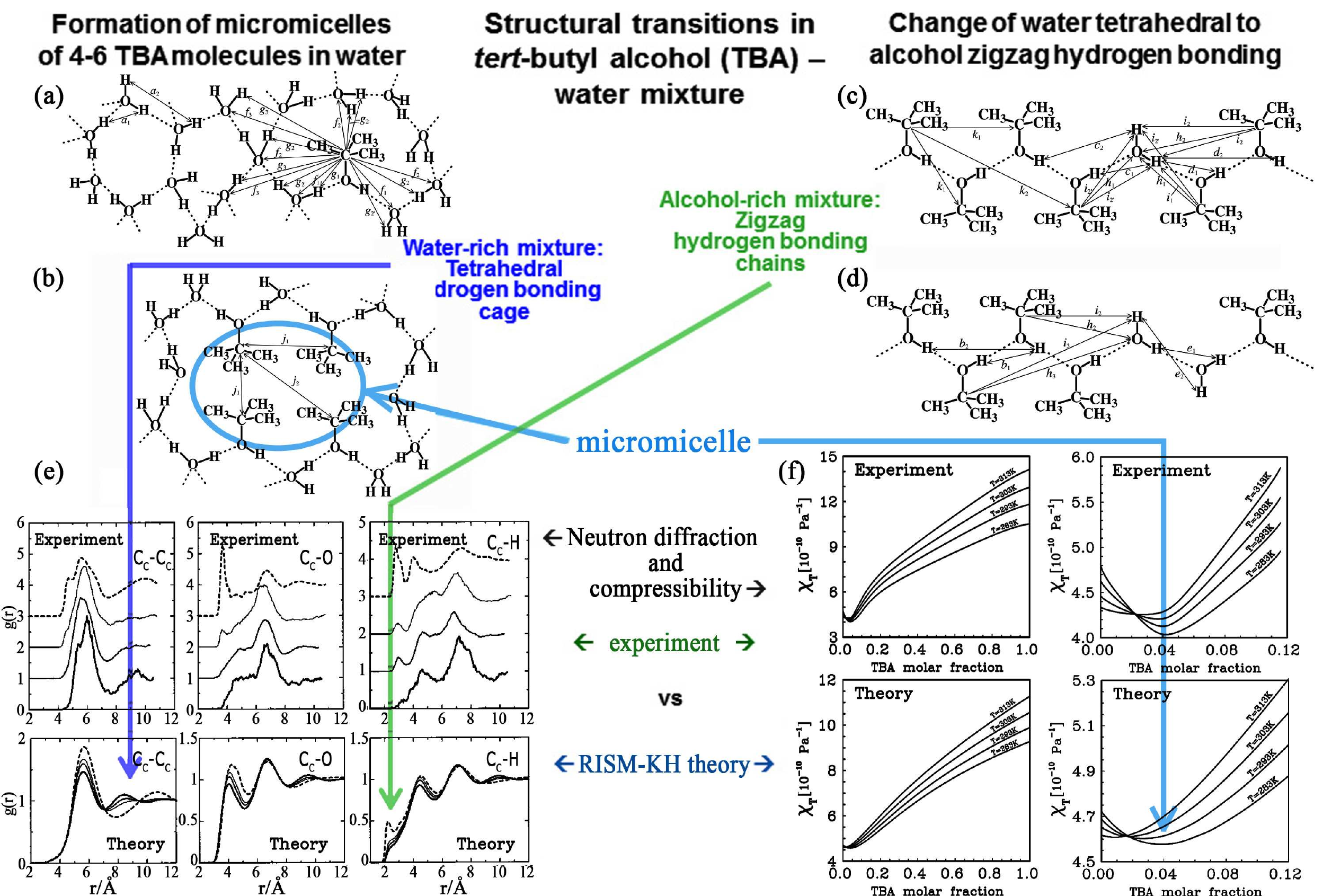}
\caption{
\label{fig:3}(Color online) Structural transitions, solvation structure and  isothermal compressibility with concentration in a mixture of water and tert-butyl alcohol (TBA) predicted by the RISM-KH molecular theory of solvation \cite{Yoshida:2002:106:5042, Omelyan:2003:2:193} versus experimental findings \cite{Bowron:1998:102:3551, Bowron:1998:93:531} (and references in \cite{Omelyan:2003:2:193}). Part (a): tert-butyl alcohol molecule at infinite dilution incorporated in a cage of water tetrahedral hydrogen bonding; part (b): micromicelle of four tert-butyl alcohol molecules in a cage of the water hydrogen bonding. Part (c): water molecules incorporated in the alcohol zigzag hydrogen bonding structure at infinite dilution of water; part (d): the same as in part (c), but at low water concentration. Part (e): solvation structure features corresponding to the water tetrahedral hydrogen bonding at low alcohol -- high water and the alcohol zigzag hydrogen bonding at high alcohol -- low water concentrations. Part (f): isosbestic point and minimum of the compressibility corresponding to the formation of TBA micromicelles in water.
}
\end{figure}

A further important example of crucial importance in chemistry and biomolecular nanosystems is the formation of nanostructures in solution driven by hydrophobic attraction. Figure~\ref{fig:3} illustrates the RISM theory predictions for the structure of the ambient mixtures of water and tert-butyl alcohol (TBA) in the whole range of concentrations \cite{Yoshida:2002:106:5042, Omelyan:2003:2:193}. TBA is a generic example of primitive surfactant with a hydrophobic head of four carbons and a hydrophilic ``tale'' represented by the hydroxyl group. The solvation structure of this system successively goes through several stages with TBA concentration in water changing from infinite dilution pure TBA: a separate TBA molecule embedded in a water tetrahedral hydrogen bonding cage at infinite dilution changes to micromicelles of four to six TBA molecules in the head-to-head arrangement incorporated in a water hydrogen bonding cage at about 4\% TBA molar fraction; then, the tetrahedral hydrogen bonding structure of water gets disrupted at about 40\% TBA molar fraction to be replaced by the zigzag hydrogen bonding structure of alcohol, with separate water molecules embedded in it at infinite dilution of water in TBA. The RISM theory predicted both the structure and thermodynamics of these mixtures, in particular, the concentration and temperature dependence of the compressibility, including the isosbestic point and minimum corresponding to the formation of micromicelles \cite{Yoshida:2002:106:5042, Omelyan:2003:2:193}, in an excellent agreement with the findings from neutron diffraction experiment \cite{Bowron:1998:102:3551, Bowron:1998:93:531} and compressibility measurements (see references in \cite{Omelyan:2003:2:193}).

\subsection{Solvation free energy of small compounds in water and octanol}
\label{sec:Water-Octanol}

The octanol-water partition coefficient characterizes hydrophobic (lipophilic) and hydrophilic properties of chemical compounds \cite{Sangster:1989:18:1111, Sangster:1997:178}. For dilute solutions it is defined as the ratio of molar concentrations of a compound in octanol and water, $P_{\textrm{O/W}}=[C_{\textrm{O}}]/[C_{\textrm{W}}]$. In practice, it is more common to use logarithm of the molar concentration ratio due to the large range of changes of the partition coefficient. As an equilibrium constant, the partition coefficient is directly related to the transfer free energy of the compound from water (polar solvent) to octanol (non-polar solvent),
\begin{equation}
  \log P_{\textrm{O/W}} = \left( \Delta\mu_{\textrm{W}} - \Delta\mu_{\textrm{O}} \right) / \left( k_{\textrm{B}}T \right) \, .
\label{eq:PartitionOW}
\end{equation}
The partition coefficient is one of the most important physico-chemical characteristics used in pharmacology, environmental studies, food industry, etc. In particular, in pharmacology, the partition coefficient is used to predict distribution of drugs within the body. It is a crucial parameter that defines drug-likeness of a compound. The partition coefficient, along with other descriptors, defines the efficiency of a drug crossing the blood brain barrier and reaching the central nervous system \cite{Waterbeemd:1998:6:151}.

Due to the practical importance, many theoretical methods have been developed to predict the partition coefficient, ranging from the phenomenological approaches based on continuum solvation models or different descriptors such as quantitative structure-activity relationship to the sophisticated approaches based on first principle calculations of the transfer free energy between different solvents, including explicit solvent MD simulations and calculations based on different solvation models with account for quantum mechanical effects (see references in \cite{Huang:2015:119:5588}). A major drawback of the continuous solvation methods is their inability to treat specific solute-solvent and solvent-solvent interactions such as hydrogen bonding, and their non-transferability (re-parametrization is required for a new solvent composition and possibly thermodynamic conditions). In principle, these difficulties can be avoided by using all-atom explicit solvent MD simulations. However, such simulations are computationally demanding and cannot be used in most practical applications such as virtual screening in rational drug design. An alternative to explicit solvent MD simulations, the 3D-RISM-KH molecular theory of solvation with the partial molar volume correction provides an excellent agreement with the experimental data for the solvation free energy in both water and 1-octanol, and so accurately predicts the octanol-water partition coefficient \cite{Huang:2015:119:5588}.

\begin{figure}[!htb]
\centering
\includegraphics[width=0.50\textwidth]{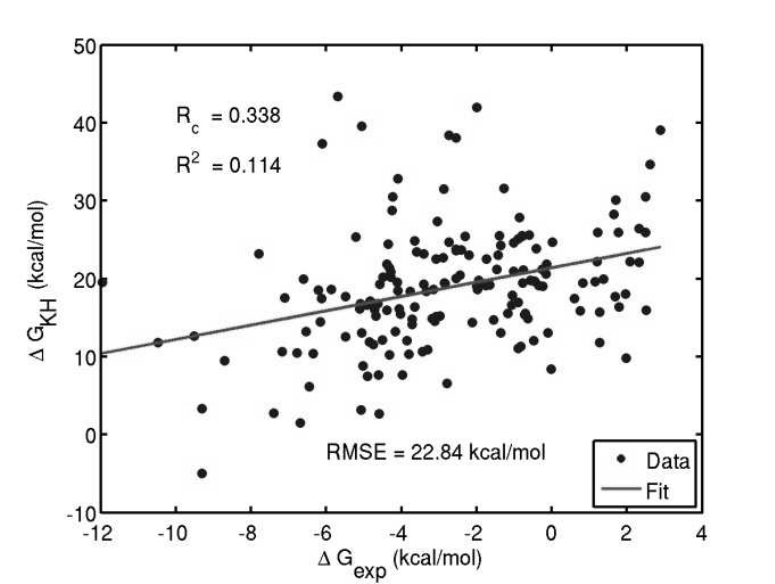}
\includegraphics[width=0.50\textwidth]{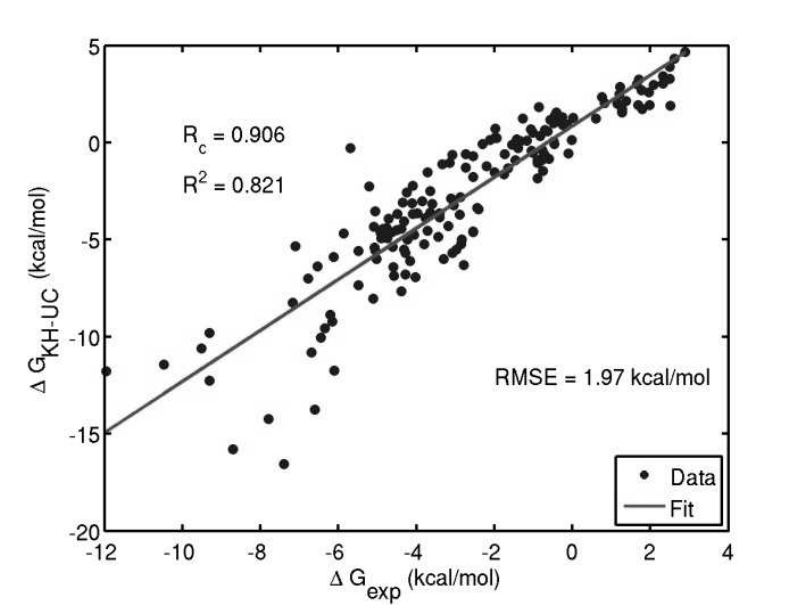}
\caption{
\label{fig:4}
Hydration free energies obtained from the KH solvation free energy functional without (top) and with (bottom) the UC term using the correlation functions from the 3D-RISM-KH theory \cite{Huang:2015:119:5588}. A comparison against experimental data for the library compounds \cite{Li:1999:103:9, Wang:2001:105:5055}.
}
\end{figure}

\begin{figure}[!htb]
\centering
\includegraphics[width=0.50\textwidth]{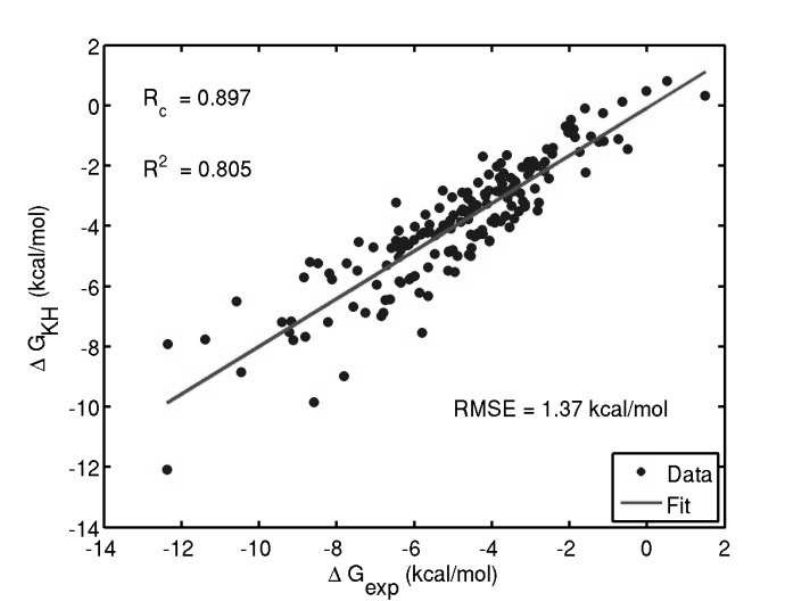}
\includegraphics[width=0.50\textwidth]{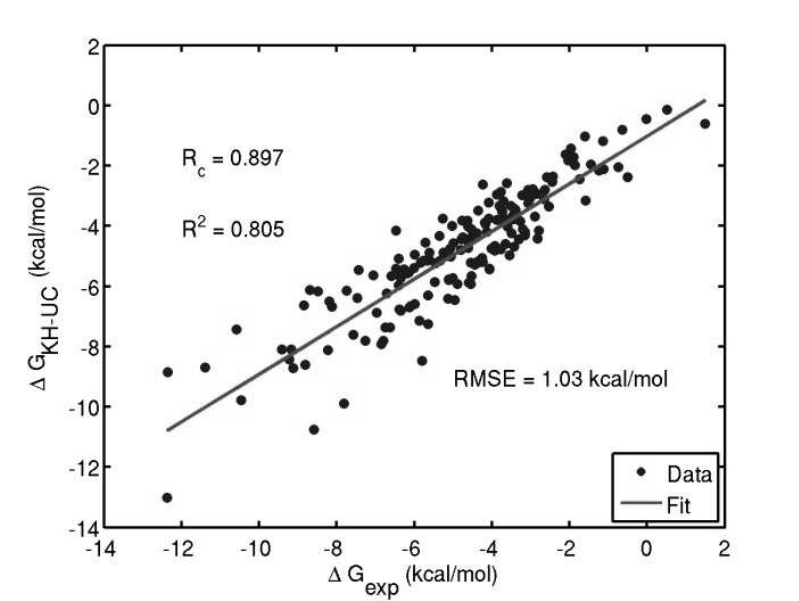}
\caption{
\label{fig:5}
Solvation free energy in octanol obtained from the KH solvation free energy functional without (top) and with (bottom) the UC term using the correlation functions from the 3D-RISM-KH theory \cite{Huang:2015:119:5588}. A comparison against experimental data for the library compounds \cite{Li:1999:103:9, Wang:2001:105:5055}.
}
\end{figure}

The 3D correlation functions of solvent around the solute compound are obtained by converging the 3D-RISM-KH integral equations \eqref{eq:3D-RISM}, \eqref{eq:3D-KH}, and then used to calculate the solvation free energy from the KH functional \eqref{eq:SFE-KH} supplemented with the UC term \eqref{eq:SFE-UC} in turn calculated from the partial molar volume \eqref{eq:PMV} using the 3D-RISM-KH correlation functions. Figures~\ref{fig:4} and \ref{fig:5} illustrate the performance of the 3D-RISM-KH theory using the KH solvation free energy functional \eqref{eq:SFE-KH} without and with the UC term \eqref{eq:SFE-UC} on solvation free energies in water and octanol \cite{Huang:2015:119:5588} for a large set of small compounds with diverse chemical groups against experimental data \cite{Li:1999:103:9, Wang:2001:105:5055}. The linear coefficients in the UC expression (19) parameterized for the solvation free energy in water and octanol calculated with the KH solvation free energy functional are displayed in Table \ref{tab:1}. The use of the UC to the KH free energy functional significantly improves the accuracy of the solvation free energy obtained from the 3D-RISM-KH theory (root mean square error (RMSE) drops from 22.84 to 1.975 kcal/mol), and thus makes it comparable to the results of explicit solvent MD simulations \cite{Palmer:2010:22:492101}. A considerably better performance of the 3D-RISM-KH theory with the KH free energy functional alone is observed in the case of solvation in octanol compared to hydration. Even in this case, the use of the UC further improves the accuracy of the results compared to experimental data (the RMSE from 1.37 down to 1.37 kcal/mol, see Table \ref{tab:1}).

\begin{table}[!htb]
\label{tab:1}
\caption{
Linear coefficients in the universal correction (UC) \eqref{eq:SFE-UC} to the Kovalenko-Hirata solvation free energy functional \eqref{eq:SFE-KH} for the two solvents \cite{Huang:2015:119:5588}. Shown is the root mean square error (RMSE) between the calculated results (with and without UC) and experimental data for the library compounds \cite{Li:1999:103:9, Wang:2001:105:5055}.
}
\vspace*{6pt}
\centering
\begin{tabular}{ | c || c | c | c | }
\hline
  \multirow{2}{*}{ Solvent }  &  \multirow{2}{*}{ $\alpha$ (kcal/mol) }  &  \multirow{2}{*}{ $\beta$ (kcal/mol) }
      &  RMSE (kcal/mol) \\
  & & & with UC (without UC) \\
\hline\hline
  Water    &  --4.58  &  0.340  &  1.975 (22.84)  \\
\hline
  Octanol  &  0.083  &  --0.939 &  1.03 (1.37) \\
\hline
\end{tabular}
\end{table}

\begin{figure}[!htb]
\centering
\includegraphics[width=0.50\textwidth]{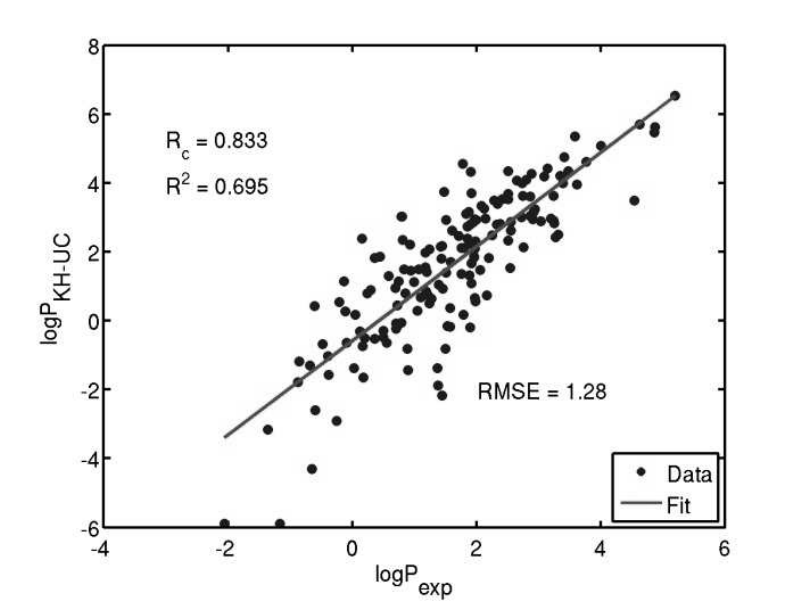}
\includegraphics[width=0.50\textwidth]{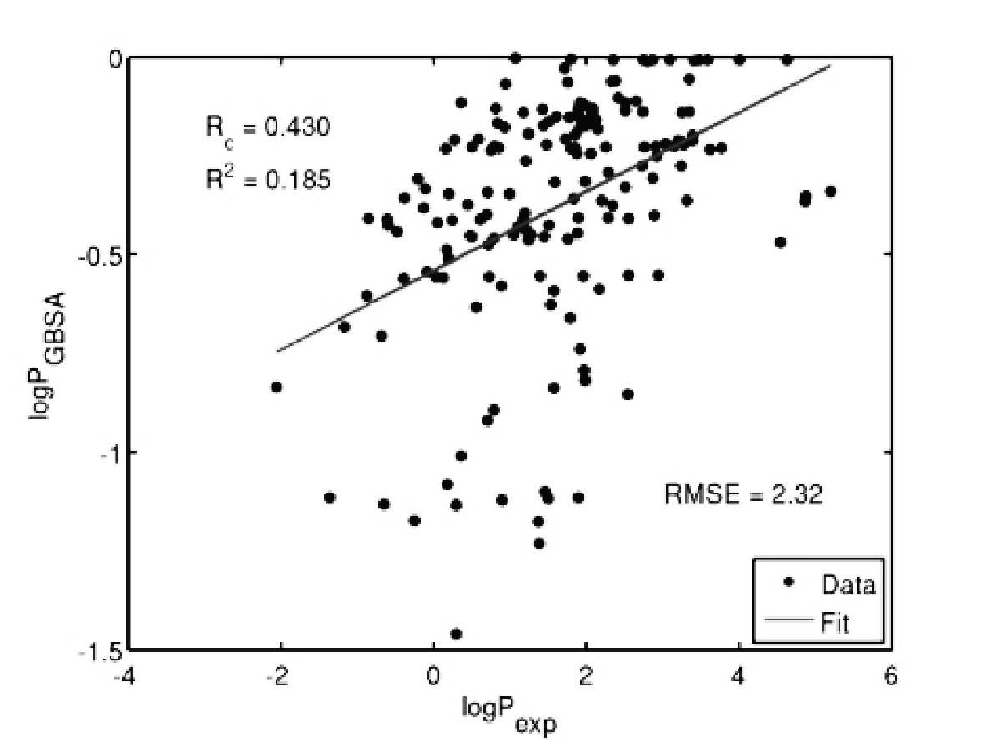}
\caption{
\label{fig:6}
Logarithm of the octanol-water partition coefficient obtained from the KH solvation free energy functional with the UC term using the correlation functions from the 3D-RISM-KH theory (top) and from the GBSA continuum solvation model (bottom) \cite{Huang:2015:119:5588}. A comparison against experimental data for the library compounds \cite{Li:1999:103:9, Wang:2001:105:5055}.
}
\end{figure}

Finally, figure~\ref{fig:6} makes a comparison of the octanol-water partition coefficient predicted from the 3D-RISM-KH theory with the KH free energy functional supplemented with the UC term for the library of compounds76 against experimental data \cite{Li:1999:103:9, Wang:2001:105:5055}. Shown for comparison are the results obtained for this library of small compounds by using the generalized Born solvent accessible surface area (GBSA) continuum solvation model widely popular in biomolecular calculations \cite{Huang:2015:119:5588}. Compared to experimental data, the prediction accuracy level of 3D-RISM-KH with UC reaches the root mean square error as low as 1.28 which is almost twice better than RMSE of 2.32 for the GBSA results.

\subsection{Effective interactions between cellulose nanoparticles in hydrogel}
\label{sec:Recalcitrance}

One representative illustration of the predictive capabilities of the 3D-RISM-KH molecular theory of solvation comes from the quest for fundamental understanding of the chemically-driven effective nanoscale forces that maintain plant cell wall structure \cite{Silveira:2013:135:19048, Stoyanov:2014:29:144, Silveira:2015:6:206}. Efficient conversion of lignocellulosic biomass to second-generation biofuels and valuable chemicals requires decomposition of resilient cell wall structure of plants. Overcoming biomass recalcitrance constitutes the most fundamental unsolved problem of plant-based green technologies \cite{Himmel:2007:315:804, Chundawat:2011:2:121}. Plants naturally evolved to withstand harsh external mechanical, thermal, chemical, and biological factors. Their secondary cell walls are composed of cellulose microfibrils embedded in a complex non-cellulosic matrix comprised mainly of hemicellulose and lignin which are responsible for the cohesive forces within the cell wall that entail structural support to plants \cite{Mortimer:2010:107:17409}. Current technological applications demand decomposition of this resilient structure in order to extract cell wall components for production of second-generation biofuels and other valuable chemical commodities. It is well known that cell wall recalcitrance varies among plant species and even within different phenotypes of the same plant. Close relationship between recalcitrance and chemical composition of a non-cellulosic matrix suggests that cell wall strength could be tuned by carefully controlling the matrix composition \cite{Himmel:2007:315:804, Mortimer:2010:107:17409, Ding:2012:338:1055, DeMartini:2013:6:898}. Full understanding of the chemical interactions within the cell walls is thus fundamental to gain control of the lignocellulosic biomass recalcitrance \cite{Chundawat:2011:2:121}.

MD simulations have been used to investigate decrystallization of cellulose \cite{Beckham:2011:115:4118, Payne:2011:2:1546}, the interactions between cellulose and non-cellulosic components of plant cell walls \cite{Lindner:2013:14:3390}, and the structure and dynamics of lignin \cite{Petridis:2011:133:20277}. However, extremely time-consuming and costly simulations are required to obtain adequate statistical sampling, addressing both solvation structure and thermodynamics of effective interactions in cell walls based on molecular forces. The 3D-RISM-KH molecular theory of solvation bridges the gap between molecular structure and effective forces on multiple length scales, and is uniquely capable of predicting the chemistry-driven effective interactions in plant cell walls \cite{Silveira:2013:135:19048, Stoyanov:2014:29:144, Silveira:2015:6:206}.

\begin{figure}[!htb]
\centering
\includegraphics[width=0.55\textwidth]{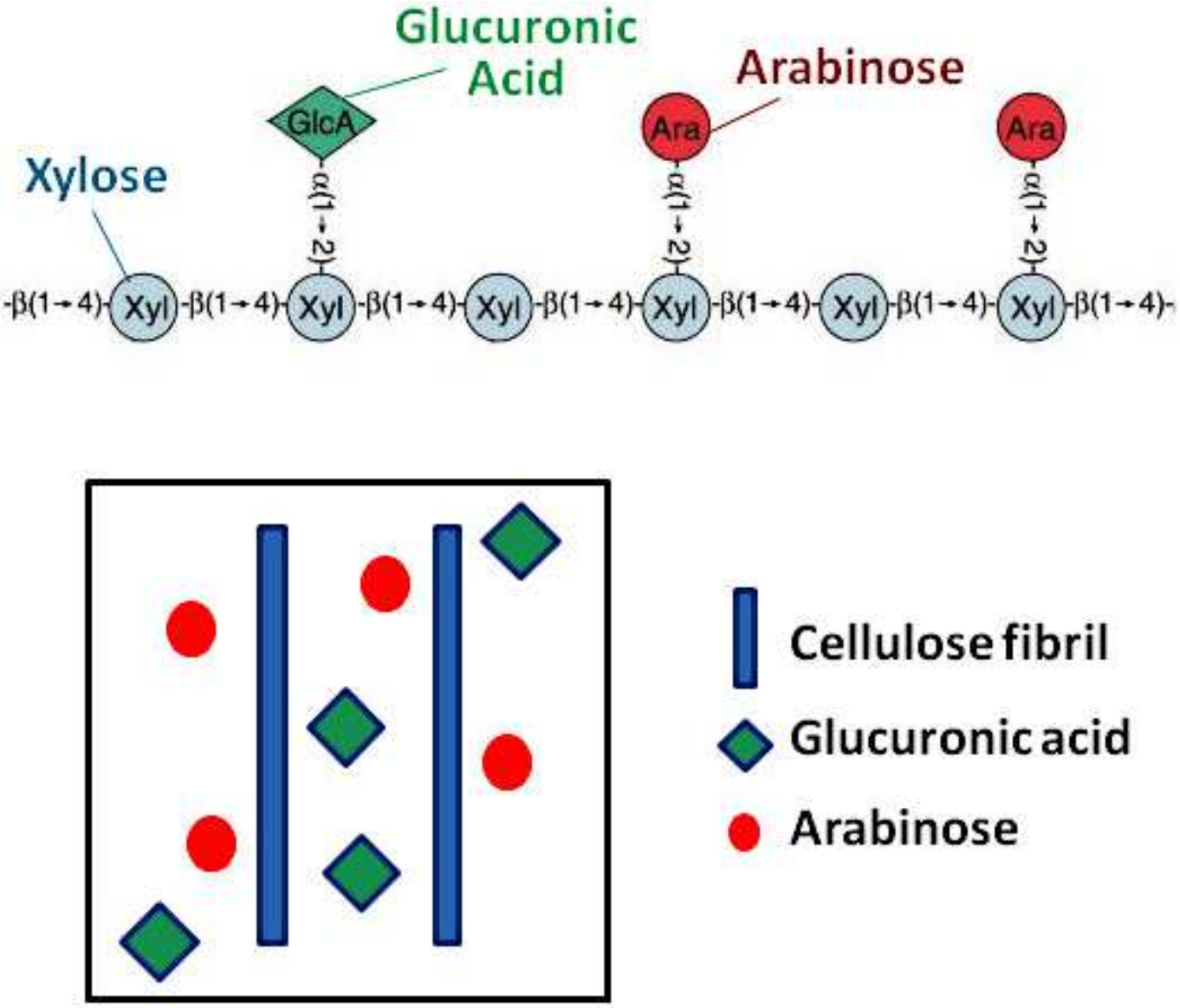}
\caption{
\label{fig:7}(Color online) Chemical structure of glucuronoarabinoxylan type hemicellulose: a xylan backbone decorated with branches of mainly arabinose and glucuronic acid in varying amount and ratio (top part). Model of plant cell walls represented with cellulose nanofibrils immersed in aqueous solution of arabinose, glucuronic acid, and glucuronate monomers at different concentrations (hemicellulose ``hydrogel'').
}
\vspace{10mm}
\end{figure}

Glucuronoarabinoxylan --- hemicellulose type most abundant in important lignocellulosic grasses used for biofuel production, such as sugar cane and corn --- consists of a xylan backbone decorated with branches of mainly arabinose and glucuronic acid whose amount and ratio substantially varies with the plant genotype \cite{Pauly:2008:54:559}. Genetic manipulation of glucuronic acid branching has been shown to significantly improve xylan extractability from cell walls without impairing plant growth \cite{Mortimer:2010:107:17409}. All the chemical changes in the xylan structure regard the branches, mainly arabinose and glucuronic acid (figure~\ref{fig:7}, top).

The modelling of structure and stability of plant cell walls consisted of the following stages \cite{Silveira:2013:135:19048}. (i) Calculation began with constructing a model of a primary cell wall represented with two 4-chain 8-glucose-long cellulose fragments, or cellulose nanocrystallites (CNs), immersed in aqueous solution of arabinose, glucuronic acid, and glucuronate monomers at different concentrations (figure~\ref{fig:7}, bottom). The I$\beta$ CNs were built by using the Cellulose-Builder\cite{Gomes:2012:33:1338} in such a way as to have both the hydrophobic and hydrophilic faces exposed to the solvent. (ii) With an input of the interaction potentials between the species in this cell wall model, the 3D-RISM-KH integral equations were solved for the 3D site correlation functions of hemicellulose monomers and water solvent (``hydrogel'') in the solvation structure of two CNs at different separations and arrangements. (iii) The ensemble-averaged 3D spatial maps of the solvation free energy density (SFED) of the hemicellulose moieties around the CNs as well as the solvation free energies were obtained for each arrangement of the CNs. (iv) Finally, the PMFs between the CNs immersed in the hemicellulose hydrogel were calculated from \eqref{eq:PMF-dir}.

\begin{figure}[!htb]
\centering
\includegraphics[width=0.990\textwidth]{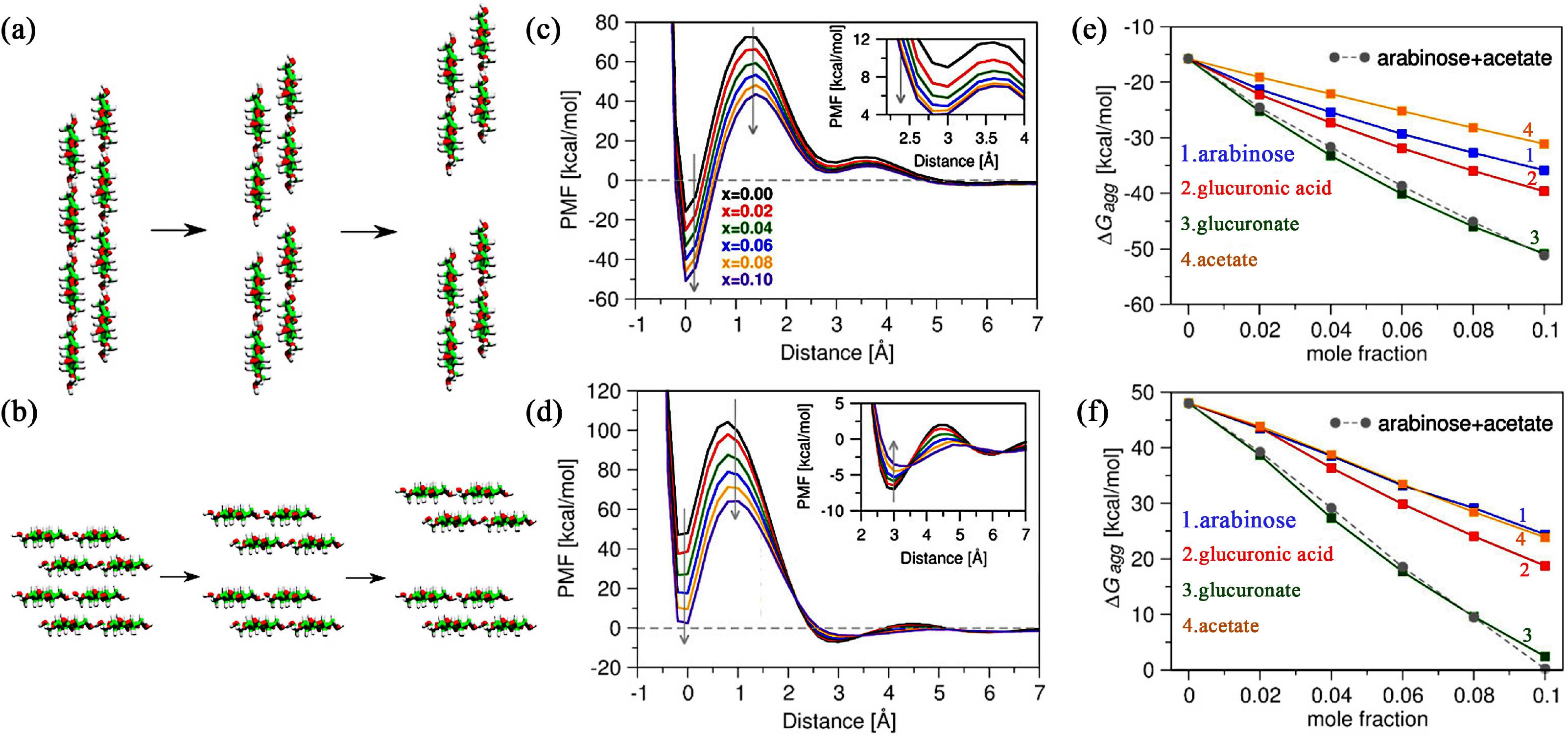}
\caption{
\label{fig:8}(Color online) Potential of mean force (PMF) and aggregation free energy $\Delta G_{\textrm{agg}}$ of cellulose nanocrystallites (CNs) in hemicellulose hydrogel. Parts (a), (b): Disaggregation of CNs by disrupting hydrophilic and hydrophobic contacts, respectively. Parts (c), (d): PMFs along disaggregation pathways a and b, respectively, at glucuronate molar fractions $x=0.0-0.1$ [legend in part (c)]. Grey arrows indicate PMF change with glucuronate concentration. Parts (e) and (f): $\Delta G_{\textrm{agg}}$ for the hydrophilic and hydrophobic contacts (a) and (b), respectively, in hemicellulose hydrogels. Grey dotted line is the sum of the arabinose and acetate curves.
}
\vspace{10mm}
\end{figure}

Figures~\ref{fig:8}~(a) and \ref{fig:8}~(b) depict the pathways of aggregation of CNs approaching each other with their hydrophilic faces (top row) and with their hydrophobic ones (bottom row), respectively. The corresponding PMFs against the separation between the CNs along these pathways are shown in figures~\ref{fig:8}~(c) and \ref{fig:8}~(d), respectively. The PMFs for disaggregation of the CNs in both the hydrophilic (part a) and hydrophobic (part b) face arrangements exhibit two well defined local minima. In both cases, the first minimum at the separation marked as $r_{\mathrm{fc}}\approx 0$~\AA{} corresponds to two aggregated CNs in direct face contact, whereas the PMF second minimum for CNs 3~\AA{} apart refers to CNs separated by a solvent layer and reaches the well depth of $-7$~kcal/mol in pure water. These results support the suggestion that CNs aggregation through hydrophilic faces is preferable to the hydrophobic faces in primary cell walls \cite{Ding:2012:338:1055}.

The stability of aggregated CNs indicates how difficult it is to disrupt their arrangement within a primary cell wall. With increase of the glucuronate concentration, the face contact aggregation free energy $\Delta G_{\textrm{agg}} = {\textrm{PMF}}(r_{\textrm{fc}})$ decreases similarly to the first maximum ${\textrm{PMF}}(r_{\textrm{bar}})$, while the disaggregation barrier $\Delta G_{\textrm{dis}} = {\textrm{PMF}}(r_{\textrm{bar}}) - {\textrm{PMF}}(r_{\textrm{fc}})$ remains the same [figures~\ref{fig:8}~(c) and \ref{fig:8}~(d)]. (The PMFs of CNs in glucuronic acid and arabinose hydrogel have similar shapes.) The larger the amount of hemicellulose, the stronger the primary plant cell wall microstructure. This is consistent with the fact that drastic structure change or absence of hemicellulose prevents plant growth due to structural collapse \cite{Pauly:2008:54:559}.

Figures~\ref{fig:8}~(e) and \ref{fig:8}~(f) show the aggregation free energies ($\Delta G_{\textrm{agg}}$) between two CNs along the hydrophilic (top) and hydrophobic face contacts (bottom). The hydrophilic face contact arrangement of the cellulose aggregate [figure~\ref{fig:8}(a)] is strongly stabilized due to interfibrillar hydrogen bonds and gives a global minimum of PMF. The global minimum for dissociation along the hydrophobic contact surface [figure~\ref{fig:8}~(b)] corresponds to the solvent-separated arrangement with the CNs $\approx 3~\AA$ apart and the well depth reaching $\approx -7$ kcal/mol in pure water.

\begin{figure}[!htb]
\centering
\includegraphics[width=0.990\textwidth]{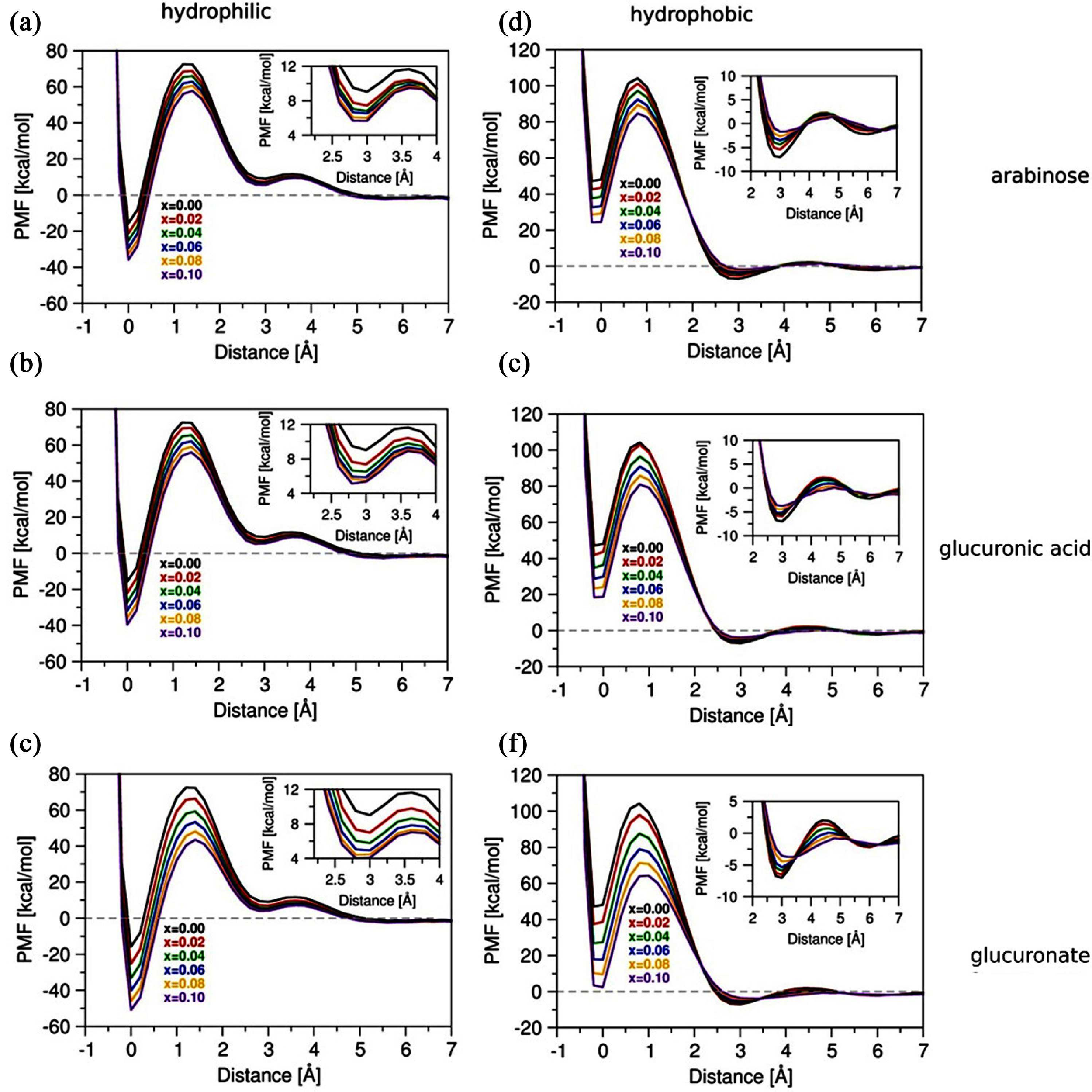}
\caption{
\label{fig:9}(Color online) Potential of mean force (PMF) for separation of the cellulose nanocrystallites along the hydrophilic and hydrophobic pathways is aqueous solution with different hemicellulose branches at different concentrations. PMFs along the hydrophilic pathway in (a) arabinose, (b) glucuronic acid, (c) glucuronate; along the hydrophobic pathway in (d) arabinose, (e) glucuronic acid, (f) glucuronate.
}
\end{figure}

Figure~\ref{fig:9} makes a comparison of the PMFs for all types of hemicellulose monomers in aqueous solution: arabinose, glucuronic acid, and glucuronate. The changes observed are quantitative. The overall features of the PMF are preserved, whereas the magnitude of the changes is strongly affected by the concentration and nature of the hemicellulose solutions.

\begin{figure}[!htb]
\centering
\includegraphics[width=0.60\textwidth]{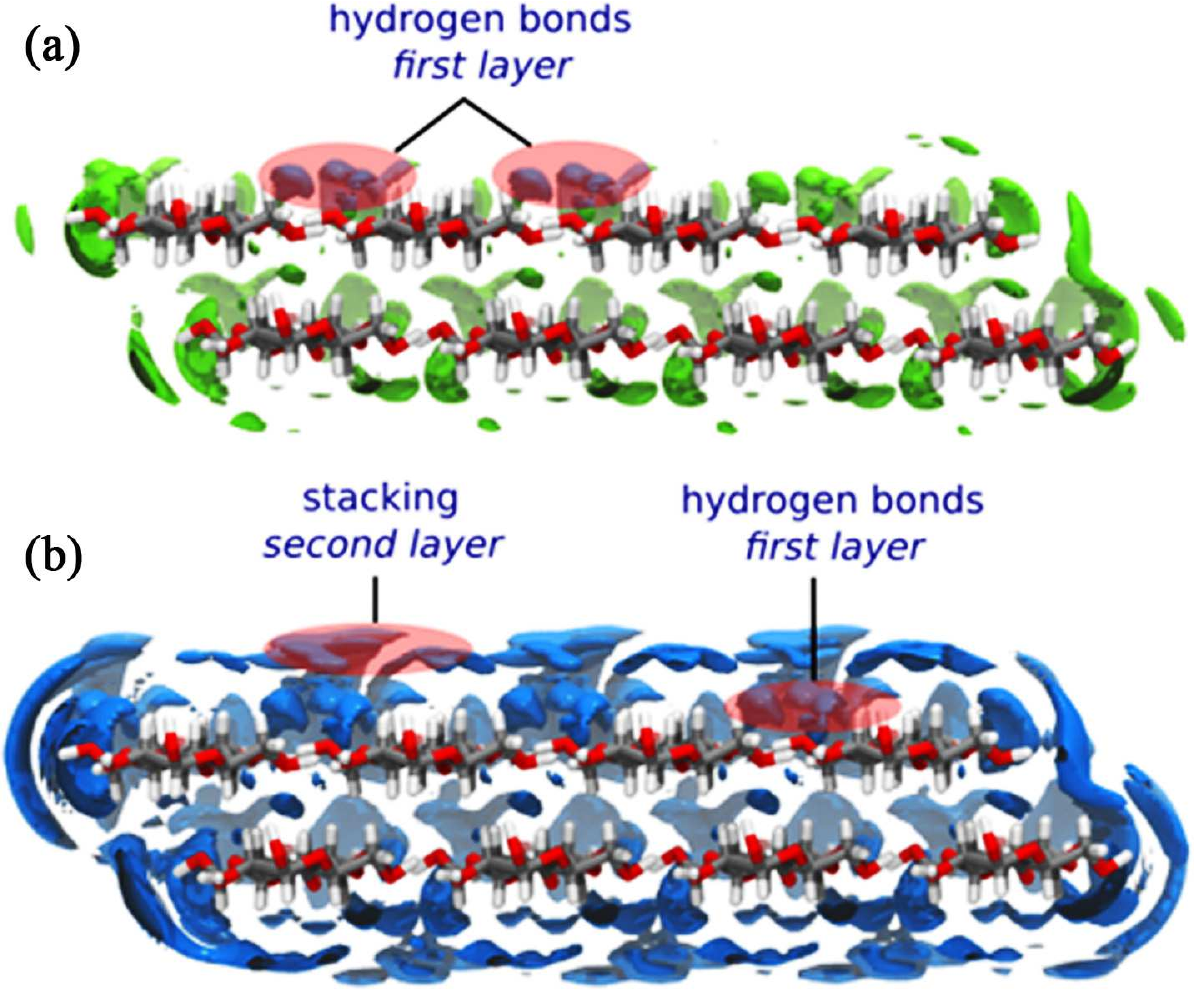}
\caption{
\label{fig:10}(Color online) Isosurfaces of the 3D solvation free energy density (3D-SFED) coming from glucuronate around the cellulose nanofibril.  Part (a): The larger negative isovalue coming from highly localized distributions corresponds to hydrogen bonding.  Part (b): The smaller negative isovalue coming from a diffuse second layer corresponding to hemicellulose-cellulose stacking.
}
\end{figure}

Finally, figure~\ref{fig:10} shows the isosurfaces of the 3D spatial maps of the solvation free energy density (3D-SFED) coming from glucuronate at the cellulose surface. The solvation free energy \eqref{eq:SFE-KH-int} of the cellulose nanofibril immersed in hemicellulose hydrogel is obtained by integrating the 3D-SFED contributions \eqref{eq:SFED-KH} from all hydrogel species over the solvation shells. The glucuronate contribution in 3D-SFED varies from large negative values for the thermodynamically most favorable arrangements of glucuronate at the cellulose surface to small negative values for less favorable arrangements. The larger negative value isosurface [figure~\ref{fig:10}~(a)] is highly localized around polar sites on the cellulose surface and indicates hydrogen bonding of hemicellulose to cellulose. The smaller negative value isosurface [figure~\ref{fig:10}~(b)] indicates a diffuse second layer of hemicellulose monomers stacking over the cellulose surface. The stacking interactions in the second layer are weaker and less specific than hydrogen bonding and are due to hydrophobic as well as enhanced inter-molecular C--H\ldots{}O interactions found in the crystal structure of cellulose \cite{Pauly:2008:54:559}. Although the stacking interactions play a considerable role, the hemicellulose-cellulose binding is controlled mainly by the site-specific H-bonds. The 3D-SFEDs for arabinose and glucuronic acid differ quantitatively, following the trends for the PMF and $\Delta G_{\textrm{agg}}$ in figures~\ref{fig:8} and \ref{fig:9}.

\section{Conclusion}
\label{sec:Conclusion}

Based on the first principles foundation of statistical mechanics and Ornstein-Zernike type integral equation theory of molecular liquids, the three-dimensional reference interaction site model with the Kovalenko-Hirata closure relation (3D-RISM-KH molecular theory of solvation) \cite{Kovalenko:1999:110:10095, Kovalenko:2000:112:10391, Kovalenko:2000:112:10403, Kovalenko:2003:169, Gusarov:2012:33:1478, Kovalenko:2013:85:159} constitutes an essential component of multiscale modelling of chemical and biomolecular nanosystems in solution. As distinct from molecular simulations which explore the phase space of a molecular system by direct sampling, the molecular theory of solvation operates with spatial distributions rather than trajectories of molecules and is based on analytical summation of the free energy diagrams which yields the solvation structure and thermodynamics in the statistical-mechanical ensemble. It provides the solvation structure by solving the integral equations for the correlation functions and then the solvation thermodynamics analytically as a single integral of a closed form in terms of the correlation functions obtained.

The 3D-RISM-KH theory yields the solvation structure in terms of 3D maps of density distribution functions of solvent interaction sites around a solute molecule with full and consistent account for the effects of chemical functionalities of all solution species. The solvation free energy and subsequent thermodynamics are then obtained at once as a simple integral of the correlation functions by performing the thermodynamic integration analytically. Moreover, the possibility of analytical differentiation of the free energy functional enables: (i) self-consistent field coupling of 3D-RISM-KH with both {\it ab initio} type and density functional theory (DFT) quantum chemistry methods for multiscale description of electronic structure in solution \cite{Kovalenko:1999:110:10095, Kovalenko:2003:169, Kovalenko:2013:85:159, Gusarov:2006:110:6083, Casanova:2007:3:458}, (ii) using 3D maps of potentials of mean force as scoring functions for molecular recognition \cite{Yoshida:2009:113:873, Genheden:2010:114:8505} and protein-ligand binding in docking protocols for fragment based drug design \cite{Yoshida:2009:113:873, Imai:2009:131:12430, Kovalenko:2011:164:101, Nikolic:2012:8:3356}, and (iii) hybrid molecular dynamics simulations performing quasidynamics of a biomolecule steered with 3D-RISM-KH mean solvation forces \cite{Miyata:2008:29:871, Luchko:2010:6:607, Omelyan:2013:39:25, Omelyan:2013:139:244106, Omelyan:2015:11:1875}.

Based on the first principles of statistical mechanics, the 3D-RISM-KH theory consistently reproduces, at the level of \textbf{fully converged} molecular simulation, the solvation structure and mean solvation forces in complex chemical and biomolecular nanosystems: both electrostatic forces (hydrogen bonding, other association, salt bridges, dielectric and Debye screening, ion localization) and nonpolar solvation effects (desolvation, hydrophobic hydration, hydrophobic interaction), as well as subtle interplays of these, such as preferential solvation, molecular recognition and ligand binding. This is very distinct from the continuum solvation schemes such as the Poisson-Boltzmann (PB) and Generalized Born (GB) models combined with the solvent accessible surface area (SASA) empirical nonpolar terms and additional volume and dispersion integral corrections, which are parameterized for hydration free energy of biomolecules but are neither really applicable to solvation structure effects in complex confined geometries nor transferable to solvent systems with cosolvent or electrolyte solutions at physiological concentrations.

For simple and complex solvents and solutions of a given composition, including buffers, salts, polymers, ligands and other cofactors at a finite concentration, the 3D-RISM-KH molecular theory of solvation properly accounts for chemical functionalities by representing in a single formalism both electrostatic and non-polar features of solvation, such as hydrogen bonding, structural solvent molecules, salt bridges, solvophobicity, and other electrochemical, associative and steric effects. For real systems, solving the 3D-RISM-KH integral equations is far less computationally expensive than running molecular simulations which should be long enough to sample relevant exchange and binding events. This enables us to handle complex systems and processes occurring on large space and long time scales, problematic and frequently not feasible for molecular simulations. The 3D-RISM-KH theory has been validated on both simple and complex associating liquids with different chemical functionalities in a wide range of thermodynamic conditions, at different solid-liquid interfaces, in soft matter, and in various environments and confinements. The 3D-RISM-KH theory offers a ``mental microscope'' capable of providing insights into structure and molecular mechanisms of formation and functioning of various chemical \cite{Kovalenko:1999:110:10095, Kovalenko:2000:112:10391, Kovalenko:2000:112:10403, Kovalenko:2003:169, Gusarov:2012:33:1478, Kovalenko:2013:85:159, Gusarov:2006:110:6083, Casanova:2007:3:458, Miyata:2008:29:871, Kaminski:2010:114:6082, Malvaldi:2009:113:3536, Kovalenko:2012:8:1508, Kovalenko:2001:349:496, Kovalenko:2002:1:381, Yoshida:2002:106:5042, Omelyan:2003:2:193, Huang:2015:119:5588, Shapovalov:2000:320:186, Stoyanov:2011:203, Fafard:2013:117:18556, Huang:2014:118:23821, Moralez:2005:127:8307, Johnson:2007:129:5735, Yamazaki:2010:11:361} and biomolecular \cite{Kovalenko:2013:85:159, Luchko:2010:6:607, Omelyan:2013:39:25, Omelyan:2013:139:244106, Omelyan:2015:11:1875, Yoshida:2006:128:12042, Yoshida:2009:113:873, Imai:2009:131:12430, Phongphanphanee:2008:130:1540, Phongphanphanee:2010:132:9782, Maruyama:2011:3:290, Yonetani:2008:128:186102, Maruyama:2010:114:6464, Harano:2001:114:9506, Imai:2001:59:512, Imai:2002:112:9469, Imai:2007:16:1927, Kovalenko:2015:22:575, Blinov:2010:98:282, Yamazaki:2008:95:4540, Blinov:2011:37:718, Genheden:2010:114:8505, Kovalenko:2011:164:101, Nikolic:2012:8:3356, Imai:2011:115:8288, Huang:2015:55:317, Stumpe:2011:115:319} systems and synthetic organic \cite{Moralez:2005:127:8307, Johnson:2007:129:5735, Yamazaki:2010:11:361} and biomass derived \cite{Silveira:2013:135:19048, Stoyanov:2014:29:144, Silveira:2015:6:206} nanomaterials.

\section*{Acknowledgements}

The work was supported by the National Institute for Nanotechnology, National Research Council of Canada, University of Alberta, Natural Sciences and Engineering Research Council of Canada, Alberta Prion Research Institute, Alberta Innovates Bio Solutions, Alberta Innovates Technology Futures, ArboraNano -- the Canadian Forest NanoProducts Network, Research Foundation of the State of S\"{a}o Paulo (FAPESP), and Brazilian Federal Agency for the Support and Evaluation of Graduate Education (CAPES). Computations were carried out on the high performance computing resources provided by the WestGrid -- Compute/Calcul Canada national advanced computing platform. The author is grateful to Dr. Nikolay Blinov, Dr. Sergey Gusarov, Dr. Stanislav Stoyanov, Dr. Rodrigo Silveira, Prof. Leonardo Costa, Prof. Munir Skaf, Prof. Tyler Luchko, and Prof. Dr. Ihor Omelyan for their passionate contributions and ongoing interest and participation in these works.


\ukrainianpart

\title{Молекулярна теорія сольватації: методологічний підсумок та ілюстрації}

\author{А.~Коваленко\refaddr{addr1,addr2}}

\addresses{
\addr{addr1}{Національний інститут нанотехнологій, 11421 Саскачеван Драйв, Едмонтон, AB, T6G 2M9, Канада}
\addr{addr2} {Факультет машинобудування, Альбертський університет, Едмонтон, AB, T6G 2G7, Канада}
}

\makeukrtitle

\begin{abstract}
\tolerance=3000%
Теорія інтегральних рівнянь, яка базується на принципах статистичної механіки, є багатообіцяючою з огляду на її місце в різномасштабній методології для розчинів хімічних та бімолекулярних наносистем. Стартуючи з міжмолекулярних потенціалів взаємодії, ця методологія використовує діаграмний аналіз вільної енергії сольватації в стистично-механічному ансамблі. Застосування відповідних умов замикання дає можливість звести нескінчений ланцюжок зв'язаних інтегральних рівнянь для багаточастинкових кореляційних функцій до системи рівнянь для дво- або трьохчастинкових кореляційних функцій. Розв'язок цих рівнянь дає результати для сольватаційної структури, точність яких є співмірною з результатами комп'ютерного моделювання, але має перевагу стосовно трактування ефектів і процесів, пов'язаних з великими відстанями та довгими часами, які якраз і створюють проблеми для комп'ютерного моделювання при явному врахуванні розчинника. Одна з версій цієї методології, теорія інтегральних рівнянь тривимірної моделі взаємодіючих силових центрів (3D-RISM) з умовами замикання Коваленка-Хірати (КН), дозволяє отримати 3D мапу кореляційних функцій, включаючи розподіл густини взаємодіючих силових центрів розчинника навколо супрамолекули, з повністю самоузгодженим врахуванням ефектів хімічної функціональності всіх складників розчину. Вільна енергія сольватації та відповідна їй термодинаміка природним чином отримується як інтеграл від 3D кореляційних функцій з тим, що термодинамічне інтегрування виконується аналітично. Аналітична форма функціоналу вільної енергії  дозволяє самоузгоджене поєднання 3D-RISM-КН теорії з методами квантової хімії при різномасштабному описі електронної структури розчину, використання 3D  мапи потенціалів середньої сили як рейтингової функції для молекулярного розпізнавання та білок-ліганд  зв'язування  в докових протоколах при фрагментарній розробці медичних препаратів, а також при гібридному молекулярно-динамічному моделюванні квазідинаміки біомолекул на основі 3D-RISM-КН потенціалів середньої сили. 3D-RISM-КН теорія була протестована як на простих, так і на складних асоційованих рідинах з різними хімічними зв'язками та в широкому діапазоні термодинамічних параметрів, на контакті з твердими та м'якими поверхнями різної природи та в різних середовищах. 3D-RISM-КН теорія є таким собі ``інтелектуальним мікроскопом'', який здатний демонструвати  структуру та механізм формування і функціонування різних хімічних та біологічних систем і наноматеріалів.

\keywords хімія розчинів, бімолекулярна сольватація, метод інтегральних рівнянь в теорії рідини, 3D-RISM-КН молекулярна теорія сольватації, сольватаційна структура і термодинаміка, потенціал середньої сили

\end{abstract}

\end{document}